\documentclass[11pt]{article}

\newcommand{\forArxiv}[1]{#1}
\newcommand{\forACM}[1]{}
\newcommand{\forSP}[1]{}
\newcommand{\forPrelim}[1]{}
\newcommand{\notforPrelim}[1]{#1}

\newif\ifstyleSP

\notforPrelim{
\usepackage{pgfplots}
\pgfplotsset{width=50ex,compat=1.9}
\usepackage{graphicx}
\usepackage{listings}
\usepackage{alltt}
\usepackage{multirow}
\usepackage{url}
\usepackage{makecell} % for \arrayrulewidth
}
\forArxiv{
\usepackage[margin=1in]{geometry}
\usepackage{authblk}
}

\notforPrelim{
\newcommand{\notes}[1]{} %
\newenvironment{code}{\begin{alltt}\small}{\end{alltt}}
\newcommand{\co}[1]{\mbox{\tt\small #1}} %code in a programming language
\newcommand{\m}[1]{$#1$} %things in math display
\newcommand\p[1]{\m{#1}}
\newcommand{\mypar}[1]{\vspace{2.5ex}\noindent{\bf #1.}\vspace{.5ex}\newline}
\newcommand{\projname}{Sec\-Algo}
}
\notes{
\newif\ifsakshamSix
\sakshamSixtrue
}%end \notes

\newif\ifsakshamSeven
\sakshamSeventrue

\begin{document}

\forACM{\fancyhead{}} % required for PLAS2019

\title{High-Level Cryptographic Abstractions
%%%%\thanks{This work was supported in part by NSF under grant CCF-1414078
%%%%distalgo and ONR under grant N000141512208.}
}
\date{}

\forSP{\author{}
\notes{
\author{
\IEEEauthorblockN{Christopher Kane, Bo Lin, Saksham Chand, Yanhong A. Liu}
\IEEEauthorblockA{Stony Brook University\\
{ckane,bolin,schand,liu@cs.stonybrook.edu}}}
}
}%end forSP

\forArxiv{
\author{Christopher Kane}
\author{Bo Lin}
\author{Saksham Chand}
\author{Scott D. Stoller}
\author{Yanhong A. Liu}
\affil{Stony Brook University\\
      \{ckane,bolin,schand,stoller,liu\}@cs.stonybrook.edu}
}

\forACM{
% using a shared affiliation, by including only the last \affiliation command, looks ridiculous and takes more space.
\author{Christopher Kane}
\affiliation{\institution{Stony Brook University}}
\email{ckane@cs.stonybrook.edu}
\author{Bo Lin}
\affiliation{\institution{Stony Brook University}}
\email{bolin@cs.stonybrook.edu}
\author{Saksham Chand}
\affiliation{\institution{Stony Brook University}}
\email{schand@cs.stonybrook.edu}
\author{Scott D. Stoller}
\affiliation{\institution{Stony Brook University}}
\email{stoller@cs.stonybrook.edu}
\author{Yanhong A. Liu}
\affiliation{\institution{Stony Brook University}}
\email{liu@cs.stonybrook.edu}
% temporarily remove \fancyhead to see what the header will look like.
\renewcommand{\shortauthors}{C.~Kane, B.~Lin, S.~Chand, S.~D.~Stoller, and Y.~A.~Liu}
}%end \forACM

\newcommand{\abstractText}{
    The interfaces exposed by commonly used cryptographic libraries are clumsy, 
    complicated, and assume an understanding of cryptographic algorithms.
    \notes{%
        %CMK: 08/13/2018 - Removed these two sentences. Merged into one 
        %paragraph.
        Correct use comes at great cost in programmer time, effort, and 
        expertise.
        Misuse, through poor choices of algorithm or configuration, occurs too
        frequently and jeopardizes the security of valuable user data.
    }%
    The challenge is to design high-level abstractions that require minimum
    knowledge and effort to use while also allowing maximum control when needed.
    
    This paper proposes such high-level abstractions consisting of simple 
    cryptographic primitives and full declarative configuration. These abstractions 
    can be implemented on top of any cryptographic library in any language. We 
    have implemented these abstractions in Python, and used them to write a wide
    variety of well-known security protocols, including Signal, Kerberos, and
    TLS.
    
    We show that programs using our abstractions are much smaller and easier to
    write than using low-level libraries, where size of 
    security protocols implemented is reduced by about a third on average. We show our implementation 
    incurs a small overhead, less than 5 microseconds for shared key operations and less than 341 microseconds ($<$ 1\%) for public key operations.
    We also show our abstractions are safe against main types 
    %the vast majority 
    of cryptographic misuse reported in the literature.
}%end \newcommand\abstractText

\forArxiv{\maketitle}

\notforPrelim{\begin{abstract}
\abstractText
\end{abstract}}

%% 2012 ACM Computing Classification System (CSS) concepts
%% Generate at http://dl.acm.org/ccs/ccs.cfm
\forACM{


}

\forACM{\maketitle}

\forPrelim{\abstractText}

\section{Introduction}

Existing cryptographic libraries are difficult to use. They require expertise
that most developers lack, and place tedious burdens on experienced developers.
The danger posed by this difficulty increases due to the proliferation of 
distributed computing, which requires that more developers make more extensive 
use of cryptographic libraries. The difficulty of using these libraries is
not new \cite{Anderson:1993, Price:1992}, but the extent of
the problem grows as computers become more deeply embedded in our
everyday lives. Mobile computing and the Internet of Things (IoT) especially
encourage the storage and transmission of sensitive data. Consumers will expect
such data to be properly secured, requiring that developers make greater use of
cryptographic libraries.

A symptom that reveals the underlying problem with existing cryptographic
libraries is the issue of cryptographic misuse---improper use of cryptographic APIs
leading to violations of security requirements~\cite[pp. 73]{Egele:2013}. A
string of papers, beginning in 2012, have documented widespread cryptographic
misuse in mobile applications \cite{Fahl:2012, Egele:2013, Balebkao:2014,
Li2014, Lazar:2014, Shuai2014, Chatzikonstantinou:2016, Ma:2016}. These works
define specific types of cryptographic misuse and build tools to detect them.

For example, Egele et al.~\cite{Egele:2013} found that 88\% of the 11,748
Android apps analyzed by their tool, CryptoLint, contained at least one
cryptographic misuse. A similar study by Ma et al.~\cite{Ma:2016} found that
99\% of the 8,640 Android apps they analyzed contained at least one
cryptographic misuse. However, detecting misuse is not sufficient to prevent all
misuse.

At the same time, higher-level cryptographic libraries have been developed, as discussed in Section~\ref{sec-related}, but they limit the choices that experts can make in using and experimenting with different cryptographic algorithms.
What are the right high-level abstractions that avoid different pitfalls?

{\bf This paper} proposes \projname{}, high-level abstractions for
cryptographic operations that aim to minimize the difficulty of using
cryptographic libraries in writing secure programs, for both non-experts and experts.  
SecAlgo provides simplest cryptographic primitives plus full declarative configurations.
It can be implemented on top of any
cryptographic library in any language. We have implemented \projname{} in
Python, and used it to write a variety of well-known security protocols,
including Signal~\cite{libsignal-javaSource}, Kerberos~\cite{RFC4120}, and TLS~\cite{RFC5246}.

We show that \projname{} reduces programmer
effort and increases clarity of programs; implementation of the abstractions incurs minimum overhead; and the abstractions prevent significant
types of cryptographic misuse.

We first describe the need for better cryptographic abstractions
%describe the need for high-level cryptographic abstractions in
in Section~\ref{sec-motive}. We define SecAlgo abstractions in Section~\ref{sec-lang}, and describe their implementation in
Section~\ref{sec-impl}. Section~\ref{sec-appl} shows application in implementing security protocols.
Section~\ref{sec-expe} presents experimental results.
Section~\ref{sec-related} discusses related work and concludes. 
Appendix~\ref{sec:misuse-prevention} presents studies of types of cryptographic misuse prevented.

\section{What are the right cryptographic abstractions?}
\label{sec-motive}

There are two reasons that better abstractions for
cryptographic operations are needed: (1) APIs for existing low-level cryptographic libraries make it
too difficult to properly use cryptographic primitives and (2)
% educating developers is not a sufficient solution to the problem.
existing high-level libraries with simplified interfaces place too many restrictions on what experts can choose, because of the limited expressive power of the abstractions used.

\notes{
The confidentiality, integrity, and authenticity of information in computer
applications are protected through the use of cryptographic primitives---algorithms
that guarantee basic cryptographic properties. These primitives are
implemented in cryptographic libraries and called through their application
program interface (API). Additionally, there are requirements for the correct
use of these primitives, e.g., required values for their arguments and
correct orders for calling them.
}

\mypar{Difficult-to-use low-level cryptographic libraries}
%\mypar{Difficulty of proper use}
%ANNIE: i added difficulty of use in the title.  
%CMK: That looks good to me
The low-level APIs provided by cryptographic libraries require many decisions,
including ones that must be coordinated, sometimes repeatedly, in order to
correctly use cryptographic primitives.
Making these decisions demands significant expertise and tedious manual effort. Even
experienced programmers are likely to make mistakes. Such mistakes will
compromise the security objective the programmer tries to achieve by using
the cryptographic primitives, as studied in the cryptographic misuse literature,
e.g., \cite{Egele:2013,Shuai2014,Chatzikonstantinou:2016,Ma:2016}.

\notes{
This difficulty of proper use is the source of cryptographic misuse.   Multiple studies show that misuse occurs in mobile applications at a concerning rate.  Appendix \ref{sec:misuse-prevention} presents results of our survey of four 
such studies .  
In summary, these studies found over 20,000 instances of types of misuse prevented by \projname.  This widespread
cryptographic misuse is the best evidence for the difficulty of using
cryptographic libraries with low-level APIs.
}

%CMK: 04/2018 - This section needed to be rewritten to reflect the shift in 
%CMK: focus from lack of knowledge to the difficulty and complexity of the
%CMK: interface as the main cause of the problems we are trying to address. I
%CMK: also tried to reduce the length of this subsection, but not sure how well
%CMK: I succeeded.
%\mypar{Educating developers is not a sufficient solution}
%
Proper use of low-level cryptographic APIs requires considerable security expertise.
So one might think that we can address the difficulty of
using these APIs by providing clear, thorough
documentation and high-quality examples of proper use.
However,
%We should, of course, strive to provide such documentation, but 
it will not suffice as a solution, because the real issue is 
the sheer number of low-level decisions
that must be made when using low-level APIs. 
%Even experienced users
%struggle to avoid mistakes when confronted with this complexity.

A right solution must directly address this complexity by protecting both
non-experts and experts from mistakes caused by carelessness, neglect, 
limited time and attention, exhaustion, and other conditions that afflict humans
confronted with unnecessary complexity.

\notes{
%CMK: 04/2018 - This is the previous version of this subsection.
Even simple uses of cryptographic APIs require considerable
security expertise as well as careful use of these primitives in order to avoid
cryptographic misuse. More sophisticated uses required to build secure
applications need even deeper security expertise as well as greater care and
discipline amidst an even greater number of tedious details.

Looking at the studies of cryptographic misuse reveals that there were already
thousands of mobile applications using cryptographic APIs. That number was
merely a lower-bound, and the true number of such mobile applications will only
have grown in the intervening years.

Given the increasing number of applications using cryptographic APIs, it is
unreasonable to expect all of their developers to include someone with the
necessary security expertise and assured care and discipline. If we wish to
prevent cryptographic misuse we must create a solution that protects these
developers from their lack of expertise and the error-prone nature of our shared
humanity when burdened with handling unnecessary details.
}

%\mypar{A design dilemma: Getting cryptographic abstractions just right} %Goldilocks
\mypar{Unduly restrictive high-level cryptographic libraries}
There are now a number of higher-level cryptographic libraries, Tink~\cite{Tink}, Libsodium~\cite{libsodium-Source},
Keyczar~\cite{KeyczarSource}, and pyca/cryptography's Fernet API~\cite{cryptography-ioSource}, that provide simplified
interfaces to reduce the tedious difficulty of using cryptographic functionality and assist non-expert users
to avoid mistakes. This is accomplished by offering a limited set of abstractions for common operations and
a restriction on the choice of algorithms, key sizes, and other configuration options.

However, expert users can find the limitations imposed by these high-level libraries restrictive and confining.
Experts may want to use or test particular combinations of algorithms in security protocols and secure applications.
They may find it difficult or impossible to do this when using existing high-level cryptographic libraries.

This issue may force expert users to return to low-level cryptographic libraries,
%though difficult for everyone to use, 
because the large number of arguments, options, and configurations
made available by low-level cryptographic libraries affords expert users their desired flexibility and 
expressive power.

\mypar{SecAlgo: High-level abstractions with full range of control}
What's needed is a way to abstract over cryptographic functionalities 
that combines the desirable features
of both low-level and high-level cryptographic libraries 
while avoiding their drawbacks. We want to
provide a simple, safe interface for all users to easily use 
%annie: change all access to use, and never use access for use
the cryptographic
functionalities they need, while giving expert users the expressive power
to configure the cryptographic functionalities they want.

\projname{} provides such high-level abstractions, requiring minimum, or zero, control by non-expert users, and allowing maximum, or full, control by expert users.
In \projname{}, the abstractions for determining
what cryptographic operation to perform are pulled apart from the abstractions used to determine how that
operation is performed. 
\begin{itemize}
\item 
\projname{} includes five primitive functions: \co{keygen} for key generation, \co{encrypt} and \co{decrypt} for public
key and shared key encryption, and \co{sign} and
\co{verify} for public key signing and MAC creation.  MAC is treated as a kind 
of signing, just as in Bellare et al.~\cite{Bellare:1995:XMN:646760.706005}. Cryptographic hash functions are also included
as a degenerate form of signing (if no key is supplied, \co{sign} returns a hash).  These are all the basic cryptographic operations that any application
might require. Every function except for \co{keygen}
takes only the data and key as arguments.

The user can optionally specify the encryption or signing algorithm to use when generating the key, otherwise a safe default algorithm is used. Non-expert users can use all the functionalities they need in a safe and simple way.
\item
\projname{} also provides a rich and expressive abstraction for declarative configuration. Expert users can 
declaratively configure all attributes used by all cryptographic algorithms to exert fine-grained control over the behavior of \projname{}'s primitive functions.
In this way, \projname{} provides the flexibility that expert users desire without reintroducing the complexity of
low-level cryptography libraries.
\end{itemize}

Other important cryptographic functionalities, such as key management functions,
% key derivation algorithms or key rotation systems, 
can be added to \projname{} by building them on top of the five primitive functions or by wrapping existing implementations of the functionalities in cryptographic libraries.

% TODO: merge with previous paragraph
% key derivation and key rotation are already mentioned above.
% Diffie-Hellman shared secret computation is mentioned in implementation section.

%\projname{} does not include abstractions for key derivation or 
%Diffie-Hellman shared secret computation. 
%It does not currently assist with 
%key storage, %management or %creation and storage
%key rotation, or %expiration, replace with a new key
%provide support for the secure deletion of secrets. 
%Adding support for these is part of our ongoing and future work.

\notes{
Given that cryptographic libraries are overly difficult to use and that this
difficulty cannot be overcome just through better education and training, it is
necessary to create new abstractions for cryptographic operations. These
abstractions can offer the cryptographic operations needed to provide
security guarantees for secure applications, while hiding the complexities
that cause cryptographic misuse. 
}

\section{High-level cryptographic abstractions} % for cryptographic primitives}
\label{sec-lang}

The abstractions in \projname{} are of two kinds: (1) high-level cryptographic
operations to provide confidentiality, integrity, and authenticity, and (2)
declarative configuration for all feasible combinations of options. 
We describe language constructs for the first kind
along with their essential arguments; additional arguments can be specified
either as optional arguments or using configuration. Properly
chosen configuration values---as determined by the current best practices
in security---are used as default.

\projname{} abstractions aim to serve both experts and non-experts. Experts
can use optional arguments and configuration to call and compose a wide range
of cryptographic operations. Non-experts can rely on the default arguments and configurations
to use safe cryptographic operations.
%CMK: 02/2018 - Maybe this is a better expression of the general approach we have
%CMK: taken to designing these abstractions.
\notes{
Our abstractions have been chosen to strike the optimal balance between
flexibility and safety. We provide application and protocol writers with the
options and choices they need to write programs easily and efficiently while
also protecting them against common forms of cryptographic misuse.
}
%Annie 5/6/18: if we are using this name, introduce it in intro, not burried here
%We call this set of abstractions and its implementation \projname{}.
%CMK: 05/08/2017: Done -- added name of project to paragraph in which we decribe our
%CMK: solution in the introduction; see line 134

\mypar{Abstractions for cryptographic operations} 
\projname{} has five basic cryptographic operations for writing security
protocols and secure applications: key generation, encryption, decryption,
signing, and verification. \projname{} provides an abstraction for each basic operation.  The abstractions allow simplest use of cryptographic algorithms---only the choice of shared vs.\ public key algorithms needs to be made for key generation, and only the data and key are needed for the other four operations.

All other choices are provided through declarative configurations.
Users can configure multiple interrelated arguments among a set of choices for each
argument, and the state configured at key generation is maintained through
subsequent, continued uses of the other four operations.
This contrasts with
operations, such as %computing a hash or
generating a random value, that return only a simple value to be used subsequently. %todo: terrible that this mixes some implementation here

We use a generic syntax in this section; each of the five operations can be implemented in any
programming language as a single API function.

\notes{
%CKM: Rewritten according to Annie's suggestion
%CMK: 12/14/2017 - Added to address Bo's comment requesting a justification for
%CMK: selecting this set of primitives.
\projname{} includes this set of abstractions because they account for the
primary uses of cryptography---confidentiality, authentication, and integrity.
In addition, these five operations are sufficient to write all but three of the
security protocols to which we have applied them. We may expand the set of basic
operations to include cryptographic hashing, shared secret computation, and 
random value generation.
%form a minimal
%sufficient set for expressing all security protocols we have found in the literature. 
%including those discussed in Section~\ref{sec-appl} and many more. 
%The set can easily be
%extended if we encounter an example requiring operations not covered by our
%abstractions.
}
%CMK: 02-2018 - Previously this said "encryption-decryption" algorithm, but that
%CMK: is too specific. Signing and verification algorithms also use keys.
%CMK: Using the generic "cryptographic algorithm instead."
%
%CMK: 06/12/2018 - Saksham suggest removing alternative terminology.
%We should just pick a single term and stick with it throughout.

\indent{\bf Key generation} is of the following form, where type \p{t} is the name of a
particular cryptographic algorithm, such as \co{AES} or \co{RSA}, or a
generic \co{shared} or \co{public}.

\begin{code}
    keygen type \p{t}
\end{code}
It generates and returns a key or pair of keys suitable for the 
cryptographic algorithm corresponding to type \p{t}.
%ANNIE: I never meant key types to be only pub and private.
%CMK: OK, I am not sure I ever realized that. I can do that, but it will require
%CMK: some changes to the implementation.
\begin{itemize}
\item
If \p{t} is the name of a specific shared-key (also called symmetric-key) algorithm, such
as \co{AES}, or is the generic \co{shared}, then \co{keygen} returns a single
value, a shared key.
%(also called secret key or symmetric key).
This key will be labeled with the name of the algorithm, \p{t}. If the
generic \co{shared} type is used, then the key is labeled with the name of the
%default
configured shared-key algorithm.  
%and the currently configured 
 
%currently configured
%Annie 5/6/18: what is that?  why not the default? need default to be safe!
%CMK: 05/08/2018 - The configuration always begins in the default state, but if user want to change
%CMK: the configuration shouldn't they expect that this will determine how keygen will
%CMK: behave.
%CMK: For example, suppose I know that I am always going to use Triple DES in my current
%CMK: program. I could pass '3DES' as the argument to every call to keygen, or I could
%CMK: instead alter the configuration "configure(shared_alg = '3DES'). Now any call to
%CMK: using the generic 'shared' as the type will create a key for '3DES' rather than
%CMK: AES.
%CMK: If not, then we will have to discuss how changes to the configuration will change
%CMK: the behavior of the other SecAlgo functions.

%Annie 5/6/18: what's nature?  representation?
%CMK: 05/08/2018 - Yes, I meant representation--what sort of value is used as the label.
%The nature of the label is implementation-dependent. 
%Annie 5/6/18 what implementation?  you mean language?
%CMK: 05/08/2018 - In part. All I meant here is that the abstractions are independent
%CMK: from any implementation of them. Of course, someone who implemented them in Java
%CMK: or C might use different values for the labels. But, there could be competing
%CMK: implementations of the same set of abstractions even within a single language, and
%CMK: they might implement these labels in different ways.
\item
If \p{t} is the name of a specific public-key (also called asymmetric-key) algorithm,
such as \co{RSA}, or is the generic \co{public}, then \co{keygen} returns a pair of
values, a private key and a public key.
Each key is labeled with the name of the algorithm, \co{t}.
%CMK: 06/12/2018 - Saksham suggests removal of the below paragraph as irrelevant to 
%CMK: cryptographic abstractions.
The two components of the pair can be retrieved by using a
simultaneous assignment (in languages that support such assignments, such as
Python) or by using two retrieval operations.

\end{itemize}

%In the case where a generic key type, \co{shared} or \co{public}, is used,
%the exact name of the algorithm can be declared as part of the configuration, 
%Annie 5/6/18: what is "the configuration"?  not defined.
%I think I see the problem that caused some my questions above too:
%the way i had written, if think or if I remember correctly,
%is that we just say the specified value or default if nothing specified,
%and then when describe configuration, 
%say they can do the "specify" declaratively.
%so the notion of "current" is bad/dangerous, 
%because that's very imperative.  we should always use "specified".
%If you start to talk about config, you'd have to define it first;
%but it is not defined until after the primitives.
%user should be able to understand
%the primitives w/o understanding the config.
%they cannot understand config anyway, or at least not good ways,
%before they understand the primitives.
%CMK: 05/08/2018 - Yes, I think that this was the source of my confusion, and this
%CMK: explanation helps clear things up for me.
%CMK: 06/12/2018 -- Removed on Saksham's advice due to poor grammar 
%or use the default; the default algorithm for each type uses the current best practice.
%if an alternative to the default choice is desired.
The size of the key can be specified as an optional argument to \co{keygen},
in an additional clause \co{size \p{s}} or otherwise declared as part of the
configuration; if the size is left unspecified the default is used.
%or otherwise use the default.

If \p{t} names a block cipher, such as AES, then a mode of 
operation~\cite{Dworkin:2001:SRB:2206250}---an algorithm for repeatedly applying
the block cipher to encrypt an arbitrary size plaintext, such as Cipher Block
Chaining (CBC)---must also be specified. Like key size, the mode of operation
can be specified as an optional argument in an additional clause \co{mode \p{m}}
or declared as a part of the configuration. If left unspecified, the default
mode of operation is used.
%Annie 5/6/18: see these two "or"s are harder to understand than necessary.
%Default configuration choices for
%algorithm and key size, decided using current best practice, are listed in
%Table~\ref{tab-config}.
%Annie 5/6/18: try not to put the table two pages later.
%CMK: 05/08/2018 - I will try, though moving around floats is one of the things
%CMK: I understand least well about latex.

{\bf Encryption and decryption} provide confidentiality, also
known as secrecy, and are of the forms below. 
\begin{code}
    encrypt text \p{txt} key \p{k}     
    decrypt text \p{txt} key \p{k}  
\end{code}
Function \co{encrypt} encrypts text \p{txt} using key \p{k} and returns the resulting encrypted text.
Function \co{decrypt} decrypts text \p{txt} using key \p{k} and returns the resulting decrypted text.
%These functions are the cryptographic operations used to provide confidentiality, also
%known as secrecy.
%CMK: 06/12/2018 - Saksham asks here, and below, whether there is special
%CMK: significance to noting that a single function is used for each operation.
%CMK: I admit, if I wrote this, I do not remember why. 
%We use a single function for each of encryption and decryption and it calls
They call the appropriate low-level library functions
as determined by the key type, size, and mode.%, the configuration, and the default.
%Annie 5/6/18: just key type k.
%config and default are just setting key type.
%I think we/you already put all info in the resulting key type,
%or I guess the word label that you introduced.
%CMK: 05/08/2018 - Yes, this is correct.
%
%ANNIE: this would make key type almost pointless.
%The key type
%determines whether a shared-key or public-key encryption-decrypting algorithm will be used,
%while the configuration determines exactly which algorithm of the given type
%will be used.

{\bf Signing and verification} provide authentication,
integrity, and non-repudiation, and are of the forms below.
%CMK: 12/5/2017 -The return type for sign, and the input and return types for
%CMK: verify both need to be made more precise. Currently, the implementation of
%CMK: sign returns a tuple (txt, sig), where sig is the signature over txt
%CMK: using key, k. Verify accepts this tuple as the first argument and returns
%CMK: a boolean: True if the sig is correct for txt given k, otherwise False.
%CMK: The implementation contains a variant that returns the original data txt
%CMK: if the verification of the signature succeeds, but this is non-standard.

%
\begin{code}
    sign text \p{txt} key \p{k}
    verify text \p{txt} sig \p{s} key \p{k}
    verify text \p{txt} key \p{k}
\end{code}% 
Function \co{sign} signs \p{txt} using key \p{k}. Signing has two modes: 
When the mode is \co{detached}, \co{sign} returns just the signature; 
when the mode is \co{combined}, \co{sign} returns the signed text.
%The form of the 
%combination, either a pair (tuple) or concatenation is implementation 
%dependent. The latter interface for sign is used in the NaCl~\cite{NaClSource} 
%and libsodium~\cite{libsodium-Source} cryptolibraries, where it is known as 
%``combined mode''. 
%The return value of \co{sign} is 
The mode of signing is determined by 
configuration, described below.  %annie: need to say which mode does what. 
The first form of function \co{verify}, for %intended to accommodate the result of 
\co{sign} used in \co{detached} mode, verifies signature \p{s} against \p{txt}
using key \p{k}, and returns true if verification succeeds and false otherwise. 
The second form of function \co{verify}, for %intended to accommodate the result of 
\co{sign} used in \co{combined} mode, verifies signed \p{txt} using key \p{k}
and returns the original text if verification succeeds and false otherwise.
%These functions provide access to cryptographic operations for authentication,
%non-repudiation, and integrity protection.
%We use a single function for both signing and signature verification. 
%%\co{sign} and \co{verify} 
%use the key type
%the configuration, and the defaults
%to determine which appropriate low-level library functions to use.
%%call the appropriate low-level library functions as determined by the key type  
%%and size.
%Annie 5/6/18: as for enc/dec.
%CMK: 05/08/2018 - Yes, changed.
%
%ANNIE: again, too coarse:
%whether shared key integrity protection (a Message Authentication Code (MAC)
%generating algorithm), or public key signature %library functions are applied
%to the data argument.
%CMK: Got it.

%
\notes{
Parameters for high-level cryptographic abstractions can be
configured declaratively.
}
%annie: write the precise syntax as I wrote in the slides,
%and a table with precise columns as in the slides too,
%so this looks like a language definition, not a list of English yet to be turned into language definition.
%CMK: 01/28/2018 - Done. Let me know how it looks.

\mypar{Declarative configuration} 
Configuration declaratively specifies values of parameters for cryptographic
operations, and is of the form below, where configuration item \p{item} is
assigned the value \p{value}.
\begin{code}
    configure \p{item} = \p{value}
\end{code}%
Table~\ref{tab-config} lists supported configuration items, their allowed
values, and default values.
%
%CMK: 02/13/2018 - Fixed table and surrounding text.
\newcommand{\captabConfig}{Configuration items, allowed values, and the default value.
%References are given for each of the allowed values where necessary. 
* for key size indicates that allowed values depend on the algorithm and backend library selected.
%Annie 5/6/18: this * would mean we need conditions for the table.
%then how could NIST just specify values as in the next row?
%CMK: 05/08/2018 - I am not sure I understand the question. What is it that would prevent NIST from
%CMK: specifying safe key sizes?
%CMK: NIST, and others, determine a key size is safe for an algorithm based on the number of operations
%CMK: required to break the key through some kind of attack, and how many operations are feasible
%CMK: given the current technology accessible to attackers. If a particular implementation of that algorithm
%CMK: did not support a safe key size, then it would no longer be safe to use that implementation.
%CMK: The listed default key size was chosen for the default algorithm AES. I suppose we may need to 
%CMK: specify a safe default key size for each of the algorithms we include.
** for key size indicates the default value that follows the NIST advice~\protect\cite{NIST:SP:800-57-1}.}
%\begin{table*}[htbp]
\begin{table*}[htb]
\forSP{\renewcommand{\arraystretch}{1.3}}
\forSP{\caption{\captabConfig}}
\centering
% for acmart/sigplan, this resizebox is needed, otherwise table is too wide.
%\resizebox{\textwidth}{!}{
%\begin{tabular}{@{}l||p{.2\textwidth}|c||p{.4\textwidth}@{}}
\begin{tabular}{@{~}l@{~\hfill}||@{~}p{\forACM{0.32}\forArxiv{0.2}\textwidth}@{~}|@{~}l@{~\hfill}||@{~}p{\forACM{0.32}\forArxiv{0.40}\textwidth}@{~}}
Item & Allowed values & Default value & References\\
\hline\hline
\co{key\_type} & \co{shared}, \co{public} & \co{shared} &  \\
\hline
\co{key\_type\_shared} & \co{AES}, \co{Blowfish}, \co{3DES}, \co{Salsa20},
\co{ChaCha20} & \co{AES} & AES~\cite{NIST197},
Blowfish~\cite{Schneier:1993:DNV:647930.740558}, 3DES~\cite{NIST:SP:800-67},
Salsa20~\cite{Salsa20-spec}, ChaCha20~\cite{ChaCha20-spec}\\
\hline
\co{key\_type\_public} & \co{RSA}, \co{DSA}, \co{ECDSA}  & \co{RSA} &
RSA~\cite{RFC8017}, DSA, ECDSA~\cite{NIST:FIPS:186-4} \\
\hline
\co{key\_size\_shared} & positive integer* & 256** & \\
\hline
\co{key\_size\_public} & positive integer* & 2048** & \\
\hline
\co{block\_cipher\_mode} & \co{CBC}, \co{CTR}, \co{CFB}, \co{EAX}, \co{GCM},
\co{CCM}, \co{SIV}, \co{OCB} & \co{GCM} & CBC, CTR,
CFB~\cite{Dworkin:2001:SRB:2206250}, EAX~\cite{BellareRogawayWagner-EAX},
GCM~\cite{NIST-GCM}, CCM~\cite{NIST-CCM}, SIV~\cite{RFC5297-SIV},
OCB~\cite{RFC7253-OCB} \\
\hline
\co{sign\_hash} & \co{SHA224}, \co{SHA256}, \co{SHA384}, \co{SHA512} &
\co{SHA256} & \cite{NIST:FIPS:180-4} \\
\hline
\co{sign\_mode} & \co{detached}, \co{combined} & \co{detached} & \\
\hline
\co{backend\_library} & \co{PyCrypto}, \co{PyNaCl}, \co{PyCryptodome},
\co{pyca/cryptography} & \co{PyCrypto} & PyCrypto~\cite{PyCrypto},
PyNaCl~\cite{PyNaCl-Source}, PyCryptodome~\cite{PyCryptodome},\hfill~
pyca/cryptography~\cite{cryptography-ioSource} \\
\hline
\end{tabular}
%}%end \resizebox
\forArxiv{\caption{\captabConfig}}
\forPrelim{\caption{\captabConfig}}
\forACM{\caption{\captabConfig}\vspacebelowfloat}
\label{tab-config}
\end{table*}
\notes{
\begin{itemize}
\item
Key type \co{shared} for \co{keygen}: %Shared Key Encryption Algorithm (
AES \cite{NIST197}, Blowfish \cite{Schneier:1993:DNV:647930.740558},
DES \cite{NIST:1999:FPD}.  The current default is AES. 
%ANNIE: is this right?
%CMK: Is what right? That the default is, or should be, AES?
%CMK: AES is the NIST recommended block-cipher
%ANNIE-open: then what is the "Random" stuff you called for keygen? shouldn't it be AES?
%CMK-DONE-10/25/2017: No. There is no AES option for key generation in PyCrypto.
%CMK: Keys for shared key ciphers are just bytestrings in PyCrypto, so any 
%CMK: random bytestring (generated by a secure rng, such as PyCrypto's own 
%CMK: Crypto.Random) will do.
%CMK: By contrast, the Java cryptography API contains a class called a
%CMK: KeyGenerator which is passed the name of the algorithm for which you wish
%CMK: to generate keys when instantiated. 

\item
Key type \co{public} for \co{keygen}: %Assymetric Encryption Algorithm 
(RSA \cite{RFC8017}.  RSA is also the current default.

\item 
key sizes: positive integers.  The current defaults are 256 bits for
shared-key algorithms, and 2048 bits for public-key algorithms.

\item 
Mode of operation, i.e., algorithm for cryptographic transformation of
data that uses a shared-key block cipher algorithm 
\cite[p. 4]{Dworkin:2001:SRB:2206250}: ECB, CBC, CTR.\\
The current default is CTR mode (Galis Counter Mode (GCM) is used if the
option for authenticated encryption is enabled).
%ANNIE-closed: The current default is?
%CMK: Sorry, not sure how I missed this one. Now, fixed.
%ANNIE-open: "or" is unclear because it allows a choice.
%CMK-DONE-10/25/2017: Removed the "or"

%\item Use of authenticated encryption \cite{} %ANNIE: what is this?
%Commenting this out, until I have better references and can explain it clearly.
\item 
Use of authenticated encryption: On, Off.
%ANNIE-open: why A and E are upper case?
%CMK-DONE-10/25/2017: No idea, I just often see it that way. Changed it.
%ANNIE: move cite to end of some sentence or phrase
"On" forces the use of GCM mode for authenticated encryption for block 
ciphers to protect the integrity of the encrypted text. The current
default is "Off" %\cite{}.
%ANNIE-open: "one of", which one?
%CMK-DONE-10/25/2017

\item 
Back-end cryptography library: for Python:
PyCrypto \cite{PyCrypto}, Charm \cite{CharmSource, charm13},
Pycryptdome \cite{PyCryptodome}. The current default is Pycropto.

\end{itemize}
}

Declarative configuration allows security experts to exert
control over the operation of high-level cryptographic abstractions 
in a clear and simple way. Proper default configuration 
values---ones that capture the best practices as determined by security
experts---are defined. This relieves developers of the burden of
making choices about security algorithms, key sizes, modes, etc. for which they
lack the relevant expertise or which are unnecessarily tedious to decide and
code at a low level.

\notes{
%ANNIE-open: most?
%CMK-DONE: I am not sure. All of the protocols that I am familiar with that use
%CMK: integrity protection, specify how the integrity protection and the 
%CMK: encryption should be combined, but this is just Denning-Sacco, TLS, and
%CMK: Kerberos, I think. The point here is further weakened by the fact that
%CMK: using one of the authenticated encryption modes would not harm anything.
%CMK: I will weaken the language here, but we may have to consider turning this
%CMK: on by default.
Note that use of authenticated encryption is "Off" by default because
protocols may want to specify their own composition of encryption and
authentication primitives to protect the confidentiality and integrity of
messages.
}

Configurations can be declared to apply globally, for particular 
sets of processes or communication channels, for particular scopes such as a
method scope, or specified as optional arguments to individual operations. 
Configurations declared for an enclosed scope override those declared for an
enclosing scope. 

%[CMK: Coarse-grained options (high, medium, low) or (high security vs. high
%efficiency) discussed here? In its own paragraph?]

\notes{
\begin{itemize}
\item
  List misuses and describe how each is inhibited when using our abstractions
  and library.
\end{itemize}
}

%CMK: 12/18/2017 - Removed material on misuse prevention from abstractions
%CMK: section and moved it to results section.

\section{Implementation} 
\label{sec-impl}
% annie-open: this section needs work, esp the code.  bo is pro bably the best
% person to read this paper, especially to give comments about the implementation
% and practical aspects.
%CMK-OPEN-10/25/2017: Agreed that we should have Bo look at this. I have shared
%CMK: the repo with him; hopefully, he will have time to take a look soon.
%CMK: I opened it just recently because I needed to make several changes to this
%CMK: section, because the implementation changed.
We have developed a prototype implementation of \projname{} as a Python 
module---a library of Python functions, one for each of the five basic
operations described in the previous section. 

%CMK: 05/13/2018 - Added passage concerning the new crypto operations. Very 
%CMK: unsure that I have picked the right location for this.
%We built the current implementation of
\projname{} is implemented on top of PyCrypto~\cite{PyCrypto} for shared-key encryption
using block ciphers (AES and Triple DES) in classical modes (CBC, CTR, CFB, OFB), message 
authentication code creation (HMAC), public key encryption (RSA), and public key
digital signing (RSA, DSA). \projname{} also provides %limited 
support 
for Diffie-Hellman key pair generation using pre-established Diffie-Hellman parameters 
defined in \cite{RFC3526, RFC5114}, but not yet abstraction for shared secret 
computation.
%However, we have also included some newer cryptographic operations, not available in PyCrypto,
%by building on top of 

In addition, \projname{} utilizes PyNaCl~\cite{PyNaCl-Source}, PyCryptodome~\cite{PyCryptodome}, and 
pyca/cryptography~\cite{cryptography-ioSource} (sometimes called cryptography.io \cite{AcarAPIUsability2017}) for authenticated encryption modes 
for block ciphers (CCM~\cite{NIST-CCM}, EAX~\cite{BellareRogawayWagner-EAX}, 
GCM~\cite{NIST-GCM}, SIV~\cite{RFC5297-SIV}, OCB~\cite{RFC7253-OCB}), safe
stream ciphers (Salsa20~\cite{Salsa20-spec}, ChaCha20~\cite{ChaCha20-spec}), key
derivation functions (HKDF~\cite{RFC5869-HKDF}), elliptic curve digital
signing~\cite{ed25519-Bernstein2012}, and elliptic curve Diffie-Hellman shared
secret computation~\cite{Curve25519-Bernstein2006}.

Some of these features (authenticated encryption modes, stream ciphers, elliptic
curve digital signing) fall under our %existing 
five abstractions. For the other
operations, 
%which are not yet incorporated into \projname{},
\projname{} provides a high-level wrapper API around a fixed library 
%annie: but you use multiple lib now.
implementation with a fixed set of parameters---choice of library implementation
or configuration is currently not provided for these operations. 
Choice of RSA as the default for public key encryption and signing is 
due to current 
incomplete support %incorporation %annie: not sure what this means
of elliptic curve cryptography. This will be 
updated in future work.

%CMK: By 'incorporated' I meant something like covered by our abstractions in the
%CMK: same way that other algorithms like AES or RSA are. Using ECC in DR, X3DH, and
%CMK: Signal requires extra work; you can't just call keygen, sign and verify.
%CMK: But, that wasn't at all clear from the context, and I didn't explain so it's
%CMK: better to replace that word with something more exact.

\notes{
%CMK: 05/13/2018 - This content got folded into the new paragraphs I added to 
%CMK: describe the newer crypto operations we added.
Our implementation method applies to using cryptographic libraries not only in Python, 
but also in other programming languages. Originally, \projname{} used only
PyCrypto~\cite{PyCrypto} as its back-end cryptographic library, but we have
since extended \projname{} to use also PyCryptodome~\cite{PyCryptodome},
pyca/cryptography~\cite{cryptography-ioSource}, and PyNaCl~\cite{PyNaCl-Source} 
for greater coverage of 
modern cryptographic algorithms (e.g., safe stream ciphers, authenticated
encryption modes, elliptic curve cryptography, etc.).
}
\notes{
The description of the implementation given in this section will rely on the
version
%annie: what version?
using PyCrypto as the back-end cryptography library.
}
%CMK: Altered paragraph to avoid unclear use of term 'version';

\mypar{Key generation} 
Implementation of \co{keygen} uses functions in the low-level cryptographic
libraries based on the type, size, and mode of operation specified. 
%to create keys material of the 
%type and size specified by the arguments passed to the function or specified
%through configuration (or, by the default type and size values, if those parameters are left empty).

\notes{
To properly implement the abstraction for key generation, any implementation
must meet the following challenges. First, the type, size and other arguments
passed to the \co{keygen} function must be checked to ensure that they represent
a safe choice. Second, key material must be generated randomly.
}
To use only safe algorithms and implementations, \projname{} checks all
arguments of \co{keygen} against whitelists for approved combinations of
algorithm types, key sizes, modes of operation, and hash functions. If the check
fails, then \co{keygen} terminates and reports an error.

\projname{} also makes sure that all key material---the bytes that form the 
shared secret
for shared keys, or the modulus and exponents that form the public and
private parts of the key pairs for RSA---is generated through calls to 
cryptographically strong pseudo-random number generators, which are usually
provided by low-level cryptographic libraries and operating systems.

\projname{} stores key material in a structure that also contains labels for
algorithm type (value of \co{key\_type}), key size (value of \co{key\_size\_shared} or \co {key\_size\_public}), block cipher mode of operation (value of \co{block\_cipher\_mode}),
public vs. private part of the key pair if \co{key\_type} is \co{public}, mode of signing (value of \co{sign\_mode}), 
%(combined mode or detached mode), 
and the name of the hash function used by 
MAC and %digital
signing algorithms (value of \co{sign\_hash}). The implementation uses Python's
\co{dict} to hold the key material and labels.

\notes{
%CMK: 12/12/2017 - Deciding what to do with this example.
For example, in the Python cryptographic library, PyCrypto, keys for shared key
ciphers are byte strings. As a consequence, there is no dedicated method for
generating new, random shared keys. Instead the crypographically-strong
random number generator included in PyCrypto is used to generate a sequence of
bytes of the given size. This key material is then collected into a dictionary
along with labels for the algorithm type, the key size, the mode of operation.

By contrast, PyCrypto represents public and private keys for the RSA public key
cipher as instances of the \co{\_RSAobj} class. The \co{RSA.generate} 
method creates a fresh private key instance. Another method is used to generate
the corresponding public key instance from the private key instance. Finally,
the key material can be extracted from each key instance using the
\co{exportkey} method. This key material is then collected together with the
labels describing the algorithm type, the key size, and the key type in a
dictionary.
}

\notes{
%CMK: 12/15/2017: Bo's answer to the included question was no, so I have removed
%CMK: this passage.
CMK - Is it helpful to have an example from another language, or should I stick
      to examples drawn only from the actual Python implementation using 
      PyCrypto (and maybe the incomplete Python implementation built on top of
      the Charm library, if any of it seems helpful)?

As a further example, when using the Java Cryptography Extension (JCE) API, a
fresh key is randomly generated when a cipher object of the given type is
created. The key material (a Java byte array) can be extracted from this object
for storage or transport.
}

\mypar{Encrypt, decrypt, sign, and verify} 
%In order to provide a simplifying abstraction
For encryption, decryption, signing, and verification, \projname{}
manages tedious low-level details so that they remain hidden.% from users.

First, \projname{} verifies that the input data is in a proper representation %formatted
for submission to the cryptographic functions of the low-level cryptographic
library.
\notes{
%BO: only "verify"? do we not try to transform the input as well?
%CMK: 08/15/2018 - Yes, when necessary; should have said that.
}
%we are building upon.
This includes making sure the input data is of the correct type and, if required
by the encryption algorithm and mode of operation, of the proper size.

For example, PyCrypto requires that plaintexts are bytes-like
objects~\cite{PyCrypto} (a bytes-like object is one that supports the Buffer
Protocol, such as \co{bytes}, \co{bytearray}, or \co{memoryview}).
%(which includes Python strings)~\cite{PyCrypto}.
\notes{
If the input data is not of the type required by the low-level crypto library,
\projname{} applies a transformation function.
}
If the input data is not of a compatible type and can not be safely converted to
the required type, then an error is signaled to the user.

Additionally, shared-key block algorithms have a block size--a fixed number of
bytes to which the algorithm can be applied. Some modes
of operation (such as CBC) require that every block of plaintext input must be
of the block size. When using these modes for messages whose length is not
divisible by the block size, \projname{} automatically pads the data using a method that 
is safe for the encryption algorithm. For example, when using a block
cipher in CBC mode, \projname{} applies the PKCS7~\cite{RFC5652}
padding algorithm, which appends 
$N$ bytes of value $N$ 
%the value of N in bytes,
%the value of N in a byte, N times,
%the value of N represented in a type, N times,
%the byte representation of value N, N times,
to pad the plaintext to a multiple of the block size, where:
\smallskip\\\indent $N = block\_size - (plaintext\_length~\%~block\_size)$\smallskip\\
Any padding is stripped from the decrypted data before it is returned.
To avoid leaking
information used by padding oracle attacks, \projname{} does not report when padding errors cause decryption failure.
%during the call to \co{encrypt}, and automatically strip that padding
%during the call to \co{decrypt}.
\notes{
%BO: maybe expand a bit here, with an example?
}

Also, \projname{} generates and handles any auxiliary values used by
the selected algorithm or mode of operation: initialization vectors, counters, and nonces.
For example, Counter (CTR) mode encryption uses a counter, which can be any function
that produces a sequence of block-size bytes values guaranteed not to repeat for a
large number of iterations.
\notes{During the encryption of a message, the block cipher and key are used to encrypt the
current counter value. The result is then XOR'd with the current block of plaintext to
produce the ciphertext. The counter is then incremented and the same process is
repeated to encrypt the next block of ciphertext.}
\projname{} follows
a standard method~\cite{Dworkin:2001:SRB:2206250} to generate the initial counter value
by using a random value for the top-half of the counter and setting the
bottom-half to 0.  The random top-half value of the counter is prepended to the ciphertext.
\notes{
An initialization vector (IV) is an unpredictable set of bytes of block size
used to randomize the encryption of the first block of plaintext for block
chaining modes of operation.
A counter is used by CTR mode encryption and is a block-size set of bytes
used to randomize the encryption of each block of plaintext.
A nonce is a single-use (per key)
value and is required for authenticated encryption modes of operation and
stream ciphers to randomize the keystream used to encrypt each byte of the
plaintext.
}
\notes{
An initialization vector (IV) is an unpredictable set of bytes of block size
used to randomize the encryption of the first block of plaintext for block
chaining modes of operation. NIST guidelines~\cite{Dworkin:2001:SRB:2206250}
mandate that the IV be unpredictable, which \projname{} accomplishes by
generating a random IV whenever one is required. The IV is then prepended to the
ciphertext to be used during decryption.

The counter used by CTR mode encryption is a block-size set of bytes
used to randomize the encryption of each block of plaintext. \projname{} follows
a standard method~\cite{Dworkin:2001:SRB:2206250} to generate the counter value
by using a random value for the top-half of the counter and setting the
bottom-half to 0. \projname{} will also initialize the counter object required
by some cryptographic libraries (such as PyCrypto). Like the IV, the random
top-half value of the counter is prepended to the ciphertext.

Finally, \projname{} generates any nonces---single use (per key)
values---that are required for authenticated encryption modes of operation and
stream ciphers to randomize the keystream used to encrypt each byte of the
plaintext.
} %end notes
\notes{
If so, 
then during encryption those values are properly generated, used to encrypt the
plaintext, and then prepended to the ciphertext. During decryption, the values
are extracted from the input and then used to decrypt the ciphertext.
}
%CMK: Think about adding an example here. Check other examples to make sure you
%CMK: are not just repeating yourself.

%CMK: Maybe add detail about application of hash function to produce digest of
%CMK: message, which digest is then submitted to the public key signing 
%CMK: and verification functions of the low-level library. This is a separate
%CMK: step a user might have to take when using a low-level library (like
%CMK: PyCrypto) directly.

For public key encryption, \projname{} uses a straightforward hybrid encryption
\cite{Cramer01designand} scheme to encrypt arbitrary amounts of plaintext data.
A single call to a public key encryption method can only encrypt a number of
bytes that is less than the public key size.  \projname{} first checks the size of the
plaintext to determine whether it can be encrypted directly using the public key
algorithm, given the key size. If not, \projname{} generates a 256-bit shared key and
encrypts the data using AES in GCM mode. The new shared key is then encrypted
with the public key algorithm. The public--key--encrypted AES key is prepended to
the AES--in--GCM--mode--encrypted data, and this concatenation is returned.

The sign and verify functions are straightforward implementations of the behavior described in Section \ref{sec-lang} using cryptographic functions specified by the configuration.

%Finally, the implementation of \co{sign} and \co{verify} takes care of mode of signing automatically.
%Implementation of \co{sign} takes the text to be signed and the key to use.  
%If \co{sign\_mode} in the key is \co{detached}, then just the signature is returned.  
%If \co{sign\_mode} is \co{combined}, then the pair of the text and the signature is returned.
%Function \co{verify} can take three or two arguments. The last argument is the key.
%If \co{sign\_mode} is \co{detached}, the implementation checks that \co{verify} 
%has three arguments, and verifies the first argument as the orignal text and the second argument as the signature using the key, and returns \co{True} if verification succeeds and \co{False} otherwise.
%If the \co{sign\_mode} is \co{combined}, then the implementation checks that \co{verify}
%has two arguments, 
%verifies the first argument as the pair of the original text and the signature using the key,
%and returns the original text from the pair if verification succeeds and False otherwise.

%annie: signature needs to have return type.  but they should have already been
%described.  here need to show the implementation, not repeating anything.

%CMK: 12/15/2017 - Not clear we need a separate paragraph for describing 
%CMK: sign and verify. There are no special challenges here that were not
%CMK: already described for the cases of encrypt and decrypt.
\notes{
\mypar{Sign and verify}
}

%CMK: 12/5/2017 - Describe the Python implementation of configure?  Kinda
%CMK: boring, but also seems necessary (the absence is noticable).
%CMK: 12/18/2017 - Nope, just boring.
\notes{
\mypar{Configuration} 

The implementation of configuration we record configuration in 
a structure that associates configuration option names with their current
values. This structure is mutable, so that option values can be changed
during program execution, and the structure persists through the execution
of the program.

When the program begins the configuration structure is initially populated with
the default configuration options.

The configuration can be altered through the \co{configure} operation, which
takes a sequence of name, value pairs as argument. If a name refers to an
existing configuration option, then the value argument is assigned as the value
of that option. Otherwise, nothing happens. Value arguments are checked against
relevant whitelists to ensure that newly configured choices of algorithm, mode
of operation, key size, etc. are safe.

The current Python implementation records configurations in a dictionary, with
option names as the keys and option values as the values. The configuration is
made persistent by writing it to a file in JSON format.
}%end \notes
%
%\mypar{Limitations}

\projname{} relies on the implementations in backend libraries to provide 
protection against side-channel attacks, 
such as timing channel mitigations built into some cryptographic libraries.
Providing additional %or inherent 
rigorous protection is future work.
%The implementation does work to avoid introducing or re-introducing vulnerabilities.

\section{Application: Implementing security protocols}
\label{sec-appl}

To demonstrate the effectiveness of \projname{}, we implemented a collection
of well-known security protocols, such as Needham-Schroeder and
others in the SPORE repository~\cite{SPORE_WEB}, as well as significant parts of
more substantial protocols: TLS version 1.2~\cite{RFC5246},
%(for which we have implemented the Handshake Protocol, Change Cipher Spec Protocol, and the Record Protocol),
Kerberos version 5~\cite{RFC4120},
%(for which we have implemented the basic handshake),
and the Signal protocol~\cite{libsignal-javaSource,
Marlinspike2013A} (including its components the Double Ratchet
protocol~\cite{DoubleRatchet-Doc} and the Extended Triple Diffie-Hellman (X3DH)
protocol~\cite{X3DH-Doc}). Table~\ref{tab-examp} lists 10 of the protocols we implemented.

\notes{The TLS-1.2 implementation includes the basic handshake, 
without client authentication, the Change Cipher Spec protocol, and the record protocol.
It includes only a limited 
selection of ciphersuites, using RSA for certificate verification and exchange of the 
premaster secret. The KRB-5 implementation includes the basic handshake, 
without client authentication, cross-realm authentication, or handling of post-dated 
or proxy tickets.}

\notes{
plus a simplified version of one of them, Denning-Sacco key distribution
protocol, to be used as a precise example.}

We implemented these protocols using \projname{} plus the DistAlgo language
\cite{Liu:2012:CED:2398857.2384645,liu2017clarity,DistAlgoSource}, an extension of
Python that provides high-level primitives for creating distributed processes, passing messages, and synchronization. The combination of
%abstractions for distributed programming and cryptographic operations
\projname{} with DistAlgo enables us to write clear, high-level implementations
of security protocols.

%CMK: 02/16/2018 - Paragraph above is rewrite of messy paragraph below.
\notes{
%We examine the implementation of a simple protocol to 
%illustrate how our abstractions enable us to write executable specifications. %annie: what's point?
To
address the orthogonal challenge of writing clear, high-level distributed
algorithms, our security protocols are implemented in the DistAlgo language 
\cite{liu2017clarity, DistAlgoSource}. 
DistAlgo provides high-level abstractions for creating distribute processes that communicate with each other by message passing and synchronization.
%that greatly simplify the implementation of distributed algorithms, of which
%security protocols are an important subclass. 
In addition, DistAlgo programs extends Python and thus can easily incorporate
%Python resources, including 
our high-level
cryptographic abstractions implemented in Python.
}

The top part of Figure~\ref{fig:ds} shows the Denning-Sacco key distribution
protocol~\cite{Denning:1981:TKD:358722.358740} (a variation on the
Needham-Schroeder public key authentication
protocol~\cite{Needham:1978:UEA:359657.359659}). There are three parties, an
initiator (A), a responder (B), and a trusted authentication server (AS). The
goal is to securely establish a new shared key (CK) known only to A and B.
A acquires certificates containing its own public key (CA) and B's
public key (CB). These certificates are passed by A to B in message 3.
%A and B acquire each other's public keys from the authentication server
With the keys contained in these certificates, A and B use public key
cryptography to protect the confidentiality and integrity of the new shared key,
and to authenticate each other.

%CMK: 02/16/2018 - Paragraph above is a rewrite of the messy paragraph below with your corrections.
\notes{
%As our example, we present in 
Figure~\ref{fig:ds} shows the uncorrected version of the simplified %annie: ??? figure says full.  also uncorrected?  so why not in caption?
%CMK: I missed this when I updated the figure. The figure now contains the full, corrected Denning-Sacco protocol (at the top)
%CMK: and the simplified, uncorrected Denning-Sacco protocol we took from Blanchet's slides.
Denning-Sacco key distribution protocol \cite{BlanchetBook09, 
Denning:1981:TKD:358722.358740} (a variation on the Needham-Schroeder Public Key
protocol \cite{Needham:1978:UEA:359657.359659}, adapted for key distribution). 
%In this protocol 
There are two participants
%annie: so why make up this new and long name?
(also called roles or parties in
descriptions of security protocols), an initiator and a responder. The goal 
%of the protocol 
is to securely establish a new shared key known only to the
participants. Public-key cryptography is used to protect both the
confidentiality and the integrity of the new shared key, and to authenticate the
participants to each other.
}

%CMK: 12/21/2017 - Need to work out how to make this figure, listing work. Maybe
%CMK: I should just write out the diagram and the associated description as a
%CMK: listing (or verbatim), rather than try to use images drawn from Denning
%CMK: and Sacco's paper.

%\begin{figure}[htbp]
\begin{figure}[htb]
\centering
\fbox{
\parbox{.9\columnwidth}{
\begin{enumerate}
    \item 
    $A \rightarrow AS : A, B$
    
    \item 
    $AS \rightarrow A : CA, CB$
    
    \item 
    $A \rightarrow B : CA, CB, \{\{ CK, T\}_{S_{A}}\}_{P_{B}}$
\end{enumerate}

Where:
\begin{itemize}
\setlength{\itemsep}{.25ex}
\item
%why quotes? %CMK: Cause I took this phrase right out of their paper.
$A$ and $B$ are users, and $AS$ is a centralized key distribution facility called an Authentication Server. 
\item
$T$ is a timestamp

\item
$P_{X}$ and $S_{X}$ denote user $X$'s public key and secret (signature) key
respectively

\item
$CA = \{A, P_{A}, T\}_{S_{AS}}$ and $CB = \{B, P_{B}, T\}_{S_{AS}}$

\item
The key $CK$ is then used for encrypting messages transmitted between $A$ and $B$.
\end{itemize}

\vspace{6pt}
\hrule %they don't have line; don't add unnecessary extras
\vspace{6pt}

\begin{enumerate}
    \item 
    $A \rightarrow B : \{\{CK\}_{S_{A}}\}_{P_{B}}$
    \item
    $B \rightarrow A : \{msg\}_{CK}$
\end{enumerate}
}
}
\caption{Top: Denning-Sacco public key distribution protocol~\protect\cite[p. 535]{Denning:1981:TKD:358722.358740}.
         Bottom: Simplified Denning-Sacco key distribution protocol~\protect\cite{BlanchetBook09}.}\forACM{\vspacebelowfloat}
\label{fig:ds}
\end{figure}
 
%annie: i couldn't make sense out of the following. you are really saying that first two msgs are not needed?  so there is only one msg left?
%CMK: Yes, the key distribution in the simplified protocol is done in one message.
A simplified version of the Denning-Sacco protocol~\cite{BlanchetBook09} is
shown in the bottom part of Figure~\ref{fig:ds}. It assumes that A and B both
already possess the other's public key. As a result, A does not need to get
certificates containing those public keys from AS, which eliminates AS and the
first two messages. The simplified protocol also does away with the timestamp
(T) associated with the new shared key.  The simplified protocol extends the 
original protocol to include sending an encrypted application message.

%[CMK: Should I just include a diagram of the protocol, or provide an English
% CMK: explanation, or both?
% Bo - Yes

Figure~\ref{fig:sds} shows the simplified Denning-Sacco protocol in \projname{}
plus DistAlgo. On line 10, message 1 of the simplified protocol is sent, and
on lines 23-24 that message is received, the encrypted shared key and signature 
are decrypted, and then the signature on the shared key is verified. Lines 11-12
and 25 illustrate the use of the new shared key to transmit encrypted messages
readable only by A and B. A message (m) is encrypted and sent on line 25,
received on line 11, and decrypted on line 12.

%caption={Simplified Denning-Sacco key distribution protocol (Uncorrected).}
%label={lst:sds}
%\begin{figure}[htbp]
\begin{figure}[htb]
\centering
\lstset{language={Python}, morekeywords={self, output, await, receive, send,
received, sent, to, from_, some, each, process, setup, run, new, start, main,
keygen, encrypt, decrypt, sign, verify}}
%\begin{minipage}{1.02\columnwidth}
%\begin{lstlisting}[xleftmargin=2ex,basicstyle={\tt \footnotesize}]
\begin{lstlisting}[basicstyle={\tt \footnotesize}]
from secalgo import *
configure(sign_mode = 'combined')

class RoleA (process):       # type A process
  def setup(skA, B, pkB):    # take in params
    pass

  def run():
    k = keygen('shared')     # new shared key
    send((1, encrypt(sign(k, skA), pkB)), to=B)
    await(some(received((2, enc_m), from_=_B)))
    m = decrypt(enc_m, k)
    output('Decrypted secret:', m)
        
class RoleB (process):       # type B process
  def setup(skB, pkA):       # take in params
    self.m = 'secret'        # set secret msg

  def run():
    await(False)

  def receive(msg=(1, enc_k), from_=A):
    k = verify(decrypt(enc_k, skB), pkA)
    if k:                    # k is not false
      send((2, encrypt(m, k)), to=A)

def main():
  skA,pkA = keygen('public') # prv,pub key of A
  skB,pkB = keygen('public') # prv,pub key of B
  B = new(RoleB, (skB, pkA)) # create B
  A = new(RoleA, (skA,B,pkB))# create A
  start(B)
  start(A)
\end{lstlisting}
%\end{minipage}
\caption{Simplified Denning-Sacco key distribution protocol.}\forACM{\vspacebelowfloat}
\label{fig:sds}
\end{figure}
%
%ANNIE: from = _R  _ is important for correctness.
%CMK: Yes, thank you.
%better number only the lines w/o empty lines
%CMK: I am not sure I know how to do that.
%CMK: Got it.
%ANNIE: I put them back.  i meant not to count empty lines, but otherwise, there
%ANNIE: is no point of not including the numbers.:) 

Each role in a protocol can be defined as a distinct process class. By extending
\co{process}, a process in DistAlgo can send messages (lines 10 and 25), handle
received messages (line 22), and \co{await} for synchronization conditions to
become true (line 11).

This example illustrates several important features of \projname{}:
\begin{enumerate}

\item Functions \co{encrypt} and \co{decrypt} can transparently provide both
shared-key and public-key cryptographic operations as determined by the key type
(shared-key on lines 12 and 25; public-key on lines 10 and 23).

\item Functions \co{encrypt}, \co{decrypt}, \co{sign}, and
\co{verify} compose smooth\-ly at a high level needing no extra effort (lines 10 and 23).

\item The return behavior of \co{sign} is controlled by the configuration 
statement on line 2, where \co{sign\_mode} is set to \p{combined}. As a result, on line 10, 
\co{sign} returns a pair of \p{k} and the 
signature over \p{k}. On line 23, \co{verify} takes that pair as first 
argument, and so returns \p{k} itself, if verification succeeds (and \p{False} 
otherwise). On Line 24, we test the return from \co{verify} to ensure that the 
verification succeeded before using the key \p{k} to encrypt the secret in 
message 2.

If \co{sign\_mode} had been set to \p{detached} on Line 2, the 
following changes to the program are required:

\begin{description}

\item Line 10 is replaced with:

\begin{code}
  send((1, encrypt((k, sign(k, skA)), pkB)), to=B)
\end{code}

where the key \p{k} is included separately in the body of message 1 because 
\co{sign} will return only the signature.

\item Lines 23-24 are replaced with:

\begin{code}
  k, sig = decrypt(enc\_k, skB)
  if verify(k, sig, pkA):
\end{code}

where the key \p{k} and the signature \p{sig} must first be retrieved from the 
encrypted text before they can be passed to \co{verify}.

\end{description}

\item The developer is relieved of any extra tasks associated with using
cryptographic operations. There is no need to generate an IV or a counter, or to
pad plaintexts, for those algorithms that require it.  All those tasks are
managed in the background.

\end{enumerate}
This last point demonstrates how \projname{} simplifies decision-making about
cryptographic operations. Even for the simplified Denning-Sacco protocol:
\begin{itemize}

\item The protocol contains three calls to \co{keygen}, two
calls each to \co{encrypt} and \co{decrypt}, and one call each to \co{sign} and
\co{verify}.

\item Those three calls to \co{keygen} contain between 9 and 18 decisions
regarding operation, algorithm, key size, mode of operation, padding, decisions
with 66 possible outcomes. 

\end{itemize}
These decision points are all occasions for a
programmer, even an experienced one, to make mistakes. \projname{}
defaults ensure that all those decisions are made safely.
%without any 
%additional input, beyond the decision between \co{public} and \co{shared} key
%types.

%[CMK: ``validation'' -- is one of those words more appropriate here].

\section{Experimental evaluation}
\label{sec-expe}

We show that \projname{} allows secure programs to be written much more easily
than using lower-level libraries and incurs a minimum overhead.
%annie: dont repeat unnecessarily, risking too much manual effort to maintain and leaving inconsistencies in paper.
%, less than 6
%microseconds for each shared key cryptographic primitive and less than 341
%microseconds ($<$ 1\%) for the more expensive public key operations. 
We also show in Appendix \ref{sec:misuse-prevention}
that \projname{} prevents seven main types of cryptographic misuse that are prevalent in mobile applications.

We compare measurements of programs that directly use \projname{} with those that use
the following lower-level libraries upon which \projname{} is built:
\begin{itemize}
\item PyCrypto:
    The most widely-used general-purpose cryptography library for 
    Python~\cite[Table 1]{AcarAPIUsability2017}.

\item pyca/cryptography:
    The second-most widely-used general purpose cryptography library for
    Python~\cite[Table 1]{AcarAPIUsability2017}.

\item PyCryptodome:
    A fork of PyCrypto extended to include newer cryptographic operations; still
    not in wide use~\cite[Table 1]{AcarAPIUsability2017}.
    
\item PyNaCl:
    The best available Python interface for Curve 25519 elliptic curve
    cryptography~\cite{AcarAPIUsability2017}.
\end{itemize}
The last three libraries provide additional operations not
available in PyCrypto.

\subsection{Code size and programming effort}

\projname{} has been used to implement over 20 security protocols, including those listed in Table~\ref{tab-examp}.
%annie: But this sentence should have been about Table 3.  For Table 4, should simply say that it lists the LOC for protocols in Table 3.
We compare implementations in \projname{} with alternative implementations that use the lower-level
cryptographic libraries PyCrypto, pyca/cryptography, PyCryptodome, and PyNaCl
directly, and with abstract specifications written for protocol verification
tools.  We also compare with implementations in Java, C\#, and Python for
NS-SK, the corrected Needham--Schroeder shared key protocol.

\newcommand{\capTabExamp}{Well-known security protocols.
  %annie: i don't see * in table:
  %* indicates the original, uncorrected version of the protocol for those with known flaws.
  %CMK: I have removed the "*"s from the table because both versions of Needham-Schroeder are
  %CMK: the corrected versions.
  }
%\begin{table}[htbp]
\begin{table}[htb]
\forSP{\renewcommand{\arraystretch}{1.3}}
\forSP{\caption{\capTabExamp}}
\centering
%\resizebox{.9\columnwidth}{!}{
\begin{tabular}{@{~}l@{~\hfill}|@{~}p{0.825\columnwidth}@{~}}
    Protocol     & Description  \\
    \hline\hline
    NS-SK        & Corrected Needham-Schroeder protocol for key distribution by key server via shared key encryption \cite{Needham:1978:UEA:359657.359659} \\ 
    \hline
    NS-PK        & Corrected Needham-Schroeder protocol for mutual authentication via public key encryption                         \cite{Needham:1978:UEA:359657.359659, Needham:1987} \\ 
    \hline
    DS           & Denning-Sacco protocol for key distribution by key server and mutual authentication via public key encryption        \cite{Denning:1981:TKD:358722.358740} \\ 
    \hline
    DS Simp      & Simplified Denning-Sacco protocol for key distribution and mutual authentication via public key encryption           \cite{BlanchetBook09} \\ 
    \hline
    DHKE-1       & Diffie-Hellman key exchange protocol with mutual authentication via public key signatures \cite{Shoup99onformal}\\ 
    \hline
    SDH          & Signed Diffie-Hellman key exchange protocol          \cite{Canetti:2001:AKP:647086.715688} \\
    \hline
    X3DH         & Extended Triple Diffie-Hellman key exchange with mutual authentication via elliptic curve public key signatures      \cite{X3DH-Doc} \\
    \hline
    DR           & Double Ratchet (aka Axolotl) encrypted message exchange protocol via shared key authenticated encryption          \cite{DoubleRatchet-Doc} \\
    \hline
    Signal       & Signal: A ratcheting forward secrecy protocol for synchronous and asynchronous messaging environments                \cite{libsignal-javaSource, Marlinspike2013A} \\
    \hline
    KRB-5        & Kerberos, version 5, protocol for key distribution by key server and mutual authentication via shared key encryption     \cite{RFC4120}                 \\
    \hline
    TLS-1.2      & Transport Layer Security (Handshake), version 1.2, for key exchange and mutual authentication via public key encryption     \cite{RFC5246}                 \\
    \hline
  \end{tabular}%}
  \forArxiv{\caption{\capTabExamp}}
  \forPrelim{\caption{\capTabExamp}}
  \forACM{\caption{\capTabExamp}\vspacebelowfloat}
  \label{tab-examp}
\end{table}

Table~\ref{tab-lines} gives the LOC (number of lines of code without comments) for the
protocols listed in Table~\ref{tab-examp}.
We use LOC as an indirect measure
of programming effort and program clarity. This is common practice in
programming language literature. 

\newcommand{\capTabLines}{LOC of protocol implementations (executable on distributed machines) in
	\projname{}+Distalgo and PyCrypto+DistAlgo, and of abstract specifications
	in the languages and tools: Scyther~\cite{ScytherSource}, AVISPA~\cite{AVISPA_Library}, ProVerif~\cite{ProVerif_Source}, Tamarin~\cite{TamarinSource}, and CryptoVerif~{CryptoVerifSource}. Our implementations of X3DH, DR, and Signal
	include 58 lines of Python code taken directly from the
	specification~\cite{DoubleRatchet-Doc}.  Empty entry means we did not find
	a corresponding specification.}
\newcommand{\tabLines}{\begin{tabular}{@{~}l||@{~}r@{~}||@{~}r@{~}||r@{~}|r@{~}|r@{~}|r@{~}|r@{~}}
Protocols & \projname+DistAlgo & PyCrypto+DistAlgo & Scyther \notes{\cite{ScytherSource}}  & AVISPA \notes{\cite{AVISPA_Library}} & ProVerif \notes{\cite{ProVerif_Source}}  & Tamarin \notes{\cite{TamarinSource}}   & CryptoVerif \notes{\cite{CryptoVerifSource}} \\
\hline\hline
NS-PK     & 47           & 96       & 36       & 55        & 107       & 109       & 116         \\
\hline
NS-SK     & 46           & 68       & 41       &           & 82        &           & 94          \\
\hline
DS        & 50           & 102      &          &           & 96        &           & 120         \\
\hline
DS Simp  & 26           & 69       &          &           &           &           &             \\
\hline
DHKE-1    & 63           & 113      & 41       &           &           &           &             \\
\hline
SDH       & 39           & 73       & 35       &           & 41        & 48        & 89          \\
\hline
X3DH      & 140          & 151      &          &           &           &           &             \\
\hline
DR        & 182          & 199      &          &           &           &           &             \\
\hline
Signal    & 321          & 349      &          &           &           &           &             \\
\hline
KRB-5     & 171          & 213      & 94       & 137       &           &           & 186         \\
\hline
TLS       & (v. 1.2) 430  & (v. 1.2) 478   & (v. 1.0) 53  & (v. 1.0) 107  & (v. 1.3) 397  & (v. 1.0) 128  &  \\
\hline
\end{tabular}
} % end \tabLines
\begin{table*}[htb]
\forSP{\renewcommand{\arraystretch}{1.3}}
\forSP{\caption{\capTabLines}}
\centering
\forArxiv{\resizebox{\textwidth}{!}{\tabLines}}
\forACM{\tabLines}
\forArxiv{\caption{\capTabLines}}
\forPrelim{\caption{\capTabLines}}
\forACM{\caption{\capTabLines}\vspacebelowfloat}
\label{tab-lines}
\end{table*}

\forArxiv{\pagebreak}
\mypar{Ease of programming using \projname{}}
The simplified function calls, automated generation of auxiliary values,
declarative configuration, and carefully selected default options in the
implementation of \projname{} result in a reduction of the number of lines required
to invoke cryptographic operations, and a simplification of those lines, when
compared to other libraries.

For example, to encrypt a string \co{pt} using AES in CBC mode with a 32-byte key
using PyCrypto, one must do the following:
\begin{code}\footnotesize
    k = Random.new().read(32)
    iv = Random.new().read(AES.blocksize)
    cipher = AES.new(k, AES.MODE_CBC, iv)
    ct = iv + cipher.encrypt(pad(pickle.dumps(pt)))
\end{code}
We can perform the same operation in \projname{} as follows, where \co{key = } is optional as in Python:
\begin{code}\footnotesize
    k = keygen('shared')
    ct = encrypt(pt, key = k)
\end{code}
We see a similar reduction in the number and complexity of lines of
code for the other cryptographic operations supported by \projname{}. In
addition, we can alter the algorithm, keysize, and mode by declaring a new
configuration, without having to alter either the call to \co{keygen} or the
call to \co{encrypt}.

Our experience is that writing protocols using \projname{} plus DistAlgo is much
easier than using other languages and libraries. For simpler protocols like the
first 6 in Table~\ref{tab-examp}, we were able to implement them in \projname{}
plus DistAlgo, with LOC shown in column 2 of Table~\ref{tab-lines}, by simply following their protocol narrations.

%Seven of our
%annie: what's "of our"?  none is "ours". 
%of the 20, not listed in table 3, right?
%need fix.  
%e.g.,Seven relatively simple protocols, not listed in Table 3,
%CMK: 03/13/2018 - There are at least three (I think it's actually 5) protocols
%CMK: that I implemented that are not listed in Table 3. Recall that we removed
%CMK: three less well-known protocols (that I implemented) to make room for 
%CMK: the Signal protocols.
Seven relatively simple protocols, not listed in Table~\ref{tab-examp},
were written by undergraduates and high-school students
who, despite having had no or little familiarity with Python and being entirely new to
\projname{} and cryptography (as well as to DistAlgo and distributed
programming), were able to complete the implementation in a couple of weeks with minimal assistance.

For X3DH, DR, and Signal protocols, we were able to easily use the core protocol
specification from \cite{DoubleRatchet-Doc}, which uses pseudocode that is simply Python code. 
We then added implementations of the
lower-level cryptographic functions, which they call ``external functions'', using other libraries for 
elliptic curve
cryptography (Curve 25519 for both signing and Diffie-Hellman) and key
derivation (HKDF).

\mypar{Comparison with using PyCrypto} % plus DistAlgo}
We also implemented the protocols in Table~\ref{tab-examp} using PyCrypto (and
PyNaCl for X3DH, DR, and Signal) plus DistAlgo. Column 3 of
Table~\ref{tab-lines} shows the LOC of these implementations. Executing these
programs produces %the same execution traces %annie: right? %CMK: 08/15/2018 - Yes.
%CMK: 08/27/2018 - Thanks to Scott for fixing this sentence.
the same calls to the underlying cryptography libraries
as those in column 2.

Table~\ref{tab-opperproto} lists the number of calls to \projname{} functions that appear in each protocol implementation. 
%As described in Section~\ref{sec-impl}, %annie: sec impl is not to show this.
%%this should be said elsewhere anyway.
%\projname{} enables
%invocation of cryptographic operations using fewer, simpler lines when compared
%to using lower-level cryptography libraries. 
Each \projname{} function call
%(other than \co{keygen}) uses 2-3 
uses 1 or more fewer lines of code compared with using PyCrypto and other lower-level libraries. 
%annie: always 2-3?
%and what about keygen?  that's most important as it is called most.
As a result, protocol implementations written
using \projname{} are shorter and simpler than those written using PyCrypto and other lower-level
libraries.

\newcommand{\capTabOpPerProto}{Number of calls to \projname{} functions in each protocol implementation.}
%\begin{table*}[htbp]
\begin{table}[htb]
\forSP{\renewcommand{\arraystretch}{1.3}}
\forSP{\caption{\capTabOpPerProto}}
\centering
%\resizebox{\columnwidth}{!}{
\begin{tabular}{@{~}l@{~\hfill}||@{~}r@{~}|@{~}r@{~}|@{~}r@{~}|r@{~}|@{\,}r@{~}||@{~}r@{~}}
Protocol & \co{keygen} & \co{encrypt} & \co{decrypt} & \co{sign} & \co{verify} & Total\\
\hline
\hline
NS-SK    & 3  & 5 & 5 & 0 & 0 & 13 \\
\hline
NS-PK    & 3  & 3 & 3 & 2 & 2 & 13 \\
\hline
DS       & 4  & 1 & 1 & 3 & 5 & 14 \\
\hline
DS Simp  & 3  & 2 & 2 & 1 & 1 & 9 \\
\hline
DHKE-1   & 7  & 0 & 0 & 4 & 4 & 15 \\
\hline
SDH      & 5  & 0 & 0 & 2 & 2 & 9 \\
\hline
X3DH     & 18 & 1 & 1 & 2 & 2 & 24 \\
\hline
DR       & 11 & 1 & 1 & 3 & 1 & 17 \\
\hline
Signal   & 29 & 2 & 2 & 5 & 3 & 41 \\
\hline
KRB-5    & 6  & 6 & 6 & 6 & 6 & 30 \\
\hline
TLS-1.2  & 14 & 3 & 3 & 4 & 3 & 27 \\
\hline
\end{tabular}
%} % resizebox
\forArxiv{\caption{\capTabOpPerProto}}
\forPrelim{\caption{\capTabOpPerProto}}
\forACM{\caption{\capTabOpPerProto}\vspacebelowfloat}
\label{tab-opperproto}
\end{table}

The average percentage difference in LOC across all implementations written
using \projname{} (shown in column 2 of Table~\ref{tab-lines}) compared to
those written using low-level libraries (shown in column 3 of
Table~\ref{tab-lines}) is 31\%, that is, using \projname{} reduces LOC of protocol
implementations by almost a third on average.

\notes{
\begin{table}[htbp]
\centering
\resizebox{55ex}{!}{
\begin{tabular}{l||c|c|c|c|c|c|c|c|c|c}
\hline
\multirow{2}{*}{Protocol} & \multicolumn{2}{c|}{\co{keygen}} & 
\multicolumn{2}{c|}{\co{encrypt}} & \multicolumn{2}{c|}{\co{decrypt}} & 
\multicolumn{2}{c|}{\co{sign}} & \multicolumn{2}{c|}{\co{verify}} \\
\cline{2-11}
& S & P & S & P & S & P & S & P & S & P \\
\hline
\hline
NS-SK & 3 & - & 5 & - & 5 & - & - & - & - & - \\
\hline
NS-PK & - & 3 & - & 3 & - & 3 & - & 2 & - & 2 \\
\hline
DS & 1 & 3 & - & 1 & - & 1 & - & 3 & - & 5 \\
\hline
DS Simp  & 1 & 2 & 1 & 1 & 1 & 1 & - & 1 & - & 1 \\
\hline
DHKE-1 & - & 7 & - & - & - & - & - & 4 & - & 4 \\
\hline
SDH & - & 5 & - & - & - & - & - & 2 & - & 2 \\
\hline
X3DH & & & & & & & & & & \\
\hline
DR & & & & & & & & & & \\
\hline
Signal & & & & & & & & & & \\
\hline
KRB-5 & 6 & - & 6 & - & 6 & - & 6 & - & 6 & - \\
\hline
TLS-1.2 & 12 & 2 & 2 & 1 & 2 & 1 & 2 & 3 & 1 & 2 \\
\hline
\end{tabular}
}
\caption{Number of calls to \projname{} functions in each protocol.}
\label{tab-opperproto}
\end{table}
}

\mypar{Comparison with abstract protocol specifications} 
\notes{
Formal specification languages are designed for writing abstract descriptions
of security protocols typically for the purpose of formal reasoning and
verification. These formal specifications rely on utmost abstractions for 
cryptographic operations, message passing among distributed processes, and
stating the desired security properties of the protocol. The aim of these formal
specifications is to give us greater insight into how these protocols operate at
an abstract level, helping protocol designers find vulnerabilities that are much
harder to spot in natural language definitions of the protocols. Many of these
specification languages are paired with tools that can automatically verify that
the protocol as specified satisfies the security properties stated in the
specification.
}
Columns 4 to 8 of Table~\ref{tab-lines} show LOC of abstract specifications for
the protocols in Table~\ref{tab-examp} in the best security protocol
specification languages (for all we could find), as written by experts in these
languages. These specifications are not executable, and are used as input to
specialized verifiers of the respective languages. Our executable \projname{}
programs are actually similar in size to the most abstract of these
specifications, as evidenced by the similar LOC. The \projname{} plus DistAlgo
implementations of all but the last 2 protocols 
%For the first 6 protocols, the simpler ones, their
%\projname{} implementation
have smaller LOC than all of the abstract specifications except for those in
Scyther (SPDL~\cite{Cremers:2012:OSV:2412048}).

%We do not yet have access to full specifications for the X3DH, Double Ratchet,
%or Signal protocols.  However, formal verification efforts for Signal and its 
%component protocols have begun \cite{libsignal-javaSource, Marlinspike2013A} and we hope to be able to
%compare the length of our implementations to those specifications soon.

%CMK: For the moment the more exacting comparison with Scyther has been removed.
\notes{
We will offer a more detailed
comparison of our protocol implementations with Scyther specifications below.
}

For the last 2 protocols, TLS and Kerberos, the most significant cause
of the larger LOC of the \projname{} plus DistAlgo implementation is
functionalities omitted from
%in the \projname{} plus DistAlgo programs but not in
the abstract specifications. For example,
the abstract specifications of TLS include only the TLS Handshake protocol,
whereas the \projname{} plus DistAlgo implementation also
includes the TLS Record protocol and the TLS ChangeCipherSpec protocol. For Kerberos, none of the abstract specifications construct tickets with actual timestamps, or use those timestamps
to validate tickets, whereas the \projname{} plus DistAlgo implementation does both.

\mypar{Comparison with using other programming languages for NS-SK}
Table~\ref{tab-lines2} compares implementations of NS-SK in C\#, Java, PyCrypto
plus Python, \projname{} plus Python, PyCrypto plus DistAlgo and \projname{}
plus DistAlgo. The implementations use different libraries for
distributed programming and cryptographic operations. These implementations were
developed by ourselves or with our supervision, and they represent our best
effort, so far, to use each language in the best way. For LOC comparison, we
formatted the programs according to the suggested style of each language.

\newcommand{\capTabLinesII}{LOC of NS-SK implementations using different languages and libraries.}
%\begin{table}[htbp]
\begin{table}[htb]
\forSP{\renewcommand{\arraystretch}{1.3}}
\forSP{\caption{\capTabLinesII}}
\centering
\begin{tabular}{@{~}l|l||r@{~}}
Language & Crypto library           & NS-SK \\
\hline\hline
%C       & Openssl \cite{OpenSSL}   & --    \\
%\hline
C\#      & .NET Cryptography \cite{dotNETCrytpo}  & 364   \\
\hline
Java     & JCA \cite{JCA}           & 351   \\
\hline
Python   & PyCrypto \cite{PyCrypto} & 217  \\
\hline
Python   & \projname{}              & 170  \\
\hline
DistAlgo & PyCrypto \cite{PyCrypto} & 68   \\
\hline
DistAlgo & \projname{}              & 46    \\
\hline
\end{tabular}
\forArxiv{\caption{\capTabLinesII}}
\forPrelim{\caption{\capTabLinesII}}
\forACM{\caption{\capTabLinesII}\vspacebelowfloat}
\label{tab-lines2}
\end{table}

The C\# and Java programs required much more effort than the Python programs,
which required much more effort than the DistAlgo programs. This is also evident
in the LOC comparison.  Our experience writing these implementations confirmed
that using high-level cryptographic and communication abstractions significantly
help reduce program size and programming effort and increase program clarity.

\subsection{Running times and overhead}
%CMK: Changes to description of Experiments

We discuss three running time experiments measuring (1) the time
taken by \projname{} functions compared with using the underlying lower-level
cryptographic library directly, (2) the time of cryptographic operations vs.\
message passing in protocols, and (3) the time of NS-SK implementations using
\projname{} vs.\ using PyCrypto on top of DistAlgo and Python, vs.
implementations in Java and C\#.

All reported running times are CPU times measured on an Intel Core i5-5250U processor of 2.70GHZ with 
16GB of DDR3L memory, running Ubuntu 17.10, DistAlgo 1.0.12, and Python 3.6.3.
PyCrypto 2.6, PyCryptdome 3.5.1, pyca/cryptography 2.2.2, and PyNaCl 1.2.1 are
used for cryptographic operations. For all experiments the Python garbage
collector was disabled. For each measurement, protocols and cryptographic
operations are run in a loop for at least one second and the CPU time is
averaged over the number of iterations in order to get an accurate estimate of
the CPU time for a single execution of that protocol or operation. Each of those
measurements is repeated at least 50 times, and the average is taken.

\mypar{Overhead of \projname{} abstractions}
We fix a configuration---an algorithm, a key size, a mode of operation, and a
padding---and measure the running time of each \projname{} primitive. We measure
the same operation written using PyCrypto directly. The measurement for using
PyCrypto directly also includes the time needed to encode input to cryptographic
functions as byte strings and the reverse for output from the cryptographic
functions. This is done because it is required by the low-level libraries.
We use the pickle library for Python for encoding and decoding.

Table~\ref{tab-time1} shows that \projname{} primitives impose small
overhead for all cryptographic primitives compared with using PyCrypto.  
The overhead is $\leq$ 4.5 microseconds for all shared key primitives,  
The overhead is $\leq$ 13.02 microseconds for all public key primitives 
except for 340.54 microseconds for \co{keygen}, but all are $<$ 1\%.

\newcommand{\capTabTimeI}{Cryptographic operations and configurations used, CPU times (in microseconds) when using PyCrypto and using \projname{}, and
time increase (in microseconds) and percentage increase from PyCrypto time to \projname{} time.}
%\begin{table*}[htb]
\begin{table*}[htb]
\forSP{\renewcommand{\arraystretch}{1.3}}
\forSP{\caption{\capTabTimeI}}
\centering
%\resizebox{\textwidth}{!}{
\begin{tabular}{@{~}l|l|l||r|r||r|r@{~}}
& Operation & \hspace{5ex}Configuration & PyCrypto  & \projname & Increase & \%  Increase \\
\hline\hline
\multirow{5}{*}{\begin{tabular}{@{}c@{}}
     Shared\\ key
\end{tabular}} 
& \co{keygen}     & AES, 256, CBC, PKCS7 & 47.36    & 49.55    & 2.19 & 4.62  \\
\cline{2-7}
& \co{encrypt}    & AES, 256, CBC, PKCS7 & 65.93    & 70.43    & 4.50 & 6.83   \\
\cline{2-7}
& \co{decrypt}    & AES, 256, CBC, PKCS7 & 14.46    & 16.71    & 2.25 & 15.56 \\
\cline{2-7}
& \co{sign}       & HMAC, 256, SHA512    & 16.9    & 18.65     & 1.75 & 10.36 \\
\cline{2-7}
& \co{verify}     & HMAC, 256, SHA512    & 17.26    & 18.84    & 1.58 & 9.15  \\
\hline
\multirow{5}{*}{\begin{tabular}{@{}c@{}}
     Public\\key 
\end{tabular}}
& \co{keygen}    & RSA, 2048    & 124,526.75 & 124.867.29       & 340.54  & 0.27 \\
\cline{2-7}
& \co{encrypt}   & RSA, 2048, OAEP      & 1,322.00   & 1,322.42   & 0.42  & 0.03 \\
\cline{2-7}
& \co{decrypt}   & RSA, 2048, OAEP      & 3100.17    & 3106.41    & 6.24  & 0.20 \\
\cline{2-7}
& \co{sign}      & RSA, 2048, PKCS1     & 2995.29    & 3008.31    & 13.02 & 0.43  \\
\cline{2-7}
& \co{verify}    & RSA, 2048, PKCS1     & 698.92     & 705.21     & 6.29  & 0.90 \\
\hline
\end{tabular}
%}%end \resizebox
\forArxiv{\caption{\capTabTimeI}}
\forPrelim{\caption{\capTabTimeI}}
\forACM{\caption{\capTabTimeI}\vspacebelowfloat}
\label{tab-time1}
\end{table*}

\notes{
%CMK: 08/20/2018 - Old table data
\begin{table*}[htbp]
\centering
%\resizebox{110ex}{!}{
\begin{tabular}{l|l|l||r|r||r|r}
& Crypto operation & Configuration & PyCrypto  & \projname & Increase & \%  Increase \\
\hline\hline
\multirow{5}{*}{\begin{tabular}{@{}c@{}}
     Shared\\ key
\end{tabular}} 
& \co{keygen}     & AES, 256, CBC, PKCS7 & 45.94    & 49.79    & 3.85 & 8.38  \\
\cline{2-7}
& \co{encrypt}    & AES, 256, CBC, PKCS7 & 65.69    & 71.58    & 5.89 & 8.97 \\
\cline{2-7}
& \co{decrypt}    & AES, 256, CBC, PKCS7 & 14.41    & 16.55    & 2.14 & 14.85 \\
\cline{2-7}
& \co{sign}       & HMAC, 256, SHA512    & 16.5    & 19.08    & 2.58 & 15.64 \\
\cline{2-7}
& \co{verify}     & HMAC, 256, SHA512    & 17.49    & 18.88    & 1.39 & 7.95 \\
\hline
\multirow{5}{*}{\begin{tabular}{@{}c@{}}
     Public\\key 
\end{tabular}}
& \co{keygen}    & RSA, 2048    & 123,705.39 & 123,644.3 & -61.09 & -0.05 \\
\cline{2-7}
& \co{encrypt}   & RSA, 2048, OAEP      & 1,324.14    & 1,332.85    & 8.71 & 0.66 \\
\cline{2-7}
& \co{decrypt}   & RSA, 2048, OAEP      & 3139.55    & 3185.41    & 45.86 & 1.46 \\
\cline{2-7}
& \co{sign}      & RSA, 2048, PKCS1     & 3006.48    & 3049.14    & 42.66 & 1.42 \\
\cline{2-7}
& \co{verify}    & RSA, 2048, PKCS1     & 756.45     & 697.92    & -58.53 & -7.74 \\
\hline
\end{tabular}%}
\caption{Cryptographic operations, configurations used, CPU times (in microseconds) in PyCrypto and in \projname{},
time increase (in microseconds) and percentage increase from PyCrypto time to \projname{} time.}
\label{tab-time1}
\end{table*}
}

Public keys used by \projname{} are 8 times
as large as those used for shared key cryptography (2048-bit 
%keys for public key cryptography 
vs. 256-bit% 
%keys for shared key cryptography
).  This means
that public key primitives 
%are more expensive
may have more varied increases in running times due to memory effect, as observed.
%annie: i think the following are not correct:
%should have larger overhead generally.
%which we observe in most cases. 
%This effect may be limited by the fact that \projname{} offers 
%more options for shared key primitives, which requires more \projname{} code
%to run during calls to shared key primitives.
%
At the same time, because public key primitives are much more expensive, the percentage increases may be much smaller, again as observed.
%Note that even though public key \co{keygen} shows a much larger increase,
%it is 40 times slower than the next most expensive operation
%(public key \co{decrypt}). 
%We also note that the range between the max and min
%values, 29,856.62 microseconds, is 87 times larger than the 340.54 microsecond
%(0.27\%) difference between the PyCrypto average running time and the \projname{}
%average running time. Public key \co{keygen} is much slower and features much
%greater variability than the other operations, which could distort the result
%even when averaged over 50 repetitions.

\notes{
%CMK: 08/20/2018 - No more negative results. No more mystery concering pub key verify
The interesting results are for public key \co{keygen}, and
\co{verify}, where we find negative overhead of -61.09 microseconds
and -58.53 respectively. 
%todo possibly for future:
%In addition, the overhead for public key \co{encrypt}
%is smaller than we expected at 8.71 microseconds.

For \co{keygen}, the range between the max and min values, 32,270 microseconds,
is 500 times larger than the -61.09 microsecond (-0.05\%) difference between the PyCrypto
average running time and the \projname{} average running time, which could distort
the result even when averaged over 50 repetitions.

We are still investigating %\co{encrypt} and 
\co{verify}, but we have noticed for the PyCrypto code the average running time is close to
its minimum running time, exceeding the minimum by only 9\%. However, its max running
time, for a few of the runs, is much larger than the average, exceeding the average
by 74\%.
\notes{
that for both PyCrypto operations the average running time is very close to the
minimum running time. However, the max running time, for a few of the runs, is
much larger than the average: it is 18\%
%1.18 times
%annie: 2 times?
%CMK: 08/13/2018 - Yes, not sure what happened here.  Sorry about this. I think 
%CMK: I was looking at the difference and not the average.
larger for \co{encrypt}, and 
%75 times 
%annie: we saw 2x, not 75x
%CMK: 03/13/2018 - Yes, I mixed this one up as well.
%1.91 times
%CMK: 08/16/2018 - Made a mistake, the max is 74% larger than the average, not 91%
91\% larger for \co{verify}. 
}
This explains the %small or 
negative
overhead of \projname{} observed for 
%these operations.  
\co{verify}
The few unusually larger running time could be due to memory behavior.
}%end notes
\notes{
%CMK: 08/10/2018 - Analysis of old data
Table~\ref{tab-time1} shows that \projname{} primitives, for the most part,
impose negligible overhead, < 5 microseconds for each primitive, compared with
direct use of PyCrypto. The public key operations appear faster in \projname{}
than direct PyCrypto. However, the difference is negligibly small, $<$ 2\%.

The interesting numbers are for the public key \co{keygen} and \co{decrypt}
where the overhead from \projname{} is 4672.6 microseconds and -22.5
microseconds respectively.

The increase in running time for public key \co{keygen} is because of the
expensive construction of a pair of private key and public key as the return
value, which copies the large (2048 bit) modulus as well as the other components
of the public and private keys. Despite the seemingly large
difference between the running time of \projname{} and PyCrypto for public
\co{keygen}, the proportional difference between the two running times is less
than 4\%. %annie: where does the 4% come from?
For public key \co{decrypt}, we are still investigating the cause, but note that
the running time of this operation is the second largest, and that the
percentage increase is less than 1\%.
}
\notes{
%CMK: 05/09/2018 - This is too long. Replacing with a condensed version.
%CMK: 05/09/2018 - Thank you to Saksham who produced the condensed version above.
Table~\ref{tab-time1} lists the results. The percentage difference for shared key operations is 
much larger than for public key operations. 
%There is a sharp divide in the overhead imposed by \projname{} abstractions over direct use of PyCrypto.
For shared key operations we see a
increase of 7.32\% to 23.53\% in the running time of the \projname{} functions. For public key operations the percentage difference for using \projname{} varies from -1.84\% to 3.85\%.
Our analysis of these results needs to address three issues: (1) we need to offer some plausible
explanation for the stark disparity in the measured difference between shared key and public key operations;
(2) we need to account for the appearance of anomalous negative cost results for three of the five public
key operations; (3) we need to explain what we actually learn about the overhead imposed by \projname{} from
these measurements.
Regarding (1), the largest contributing factor is the difference in the efficiency of the underlying cryptographic
operations. Public key (RSA) operations are orders of magnitude slower than corresponding symmtric key operations.
Our measurements for PyCrypto indicate that shared key encryption is about 20 times faster than public key
encyption, and shared key decryption is more than 200 times faster than public key decryption. Even if the
overhead of \projname{} were the same for shared key and public key operations, this difference in the
efficiency of the underlying operations would adequately explain why that overhead account for a much greater
proportion of the total running time of the \projname{} functions for shared key operations. 
Additionally, the overhead cannot be the same between shared key and public key functions in \projname{}. 
Shared key operations have a larger number of configuration options, and so there is more code---in the
form of control logic, value lookups, and branching---between the \projname{} function call and the 
performance of the cryptographic operation for shared key functions.

Regarding (2), we have just seen that the overhead of \projname{} for public key operations can be no larger than the
overhead for shared key operations. The average difference in the running time between the PyCrypto and \projname{}
shared key functions is 3.42 microseconds. The range between the min and max observed running times for public key
operations is at least ten times greater than this. The most likely explanation for the anomalous results observed for
public key operations is that the \projname{} overhead was overwhelmed by the much larger variation in the running time
of the underlying cryptographic operations making it difficult to find the \projname{} overhead in the results. The
measurements would need to be repeated a much larger number of times in order to smooth out this variation and give us
a chance of measuring the overhead imposed by \projname{}. Preliminary investigations have taken a much more careful
look at several of the public key operations, averaging over a much larger number of measurements, and they currently
indicate that the \projname{} overhead accounts for less than 1\% of the running time of our public key functions.

Finally we come to issue (3). The most plausible judgment we can make about the \projname{} overhead, based on this set
of measurements, is that it varies between 2 and 4 microseconds. For public key methods, this would mean that use of
\projname{} imposes a reasonable cost of less than 1\%. Unfortunately, this extra time is a much more significant cost
for shared key methods. Even the lowest is notable at 7\%. We are investigating optimization strategies to reduce this
cost.
}
%CMK: 05/08/2017 - This description no longer applies, so I have commented it out (My first inclination was to
%CMK: delete it, or just rewrite it, but I remembered not to do that.
\notes{
For several of the shared key
operations it appears that the \projname{} operations run faster than the corresponding
PyCrypto operations. However, the differences are on the order of 10's of microseconds
and are probably due to limitations on the precision of our running time measurements. A larger
and odder discrepancy concerns the \co{keygen} operation for public keys. The difference is still
small, but given how slow public key generation we should have observed a lower running time for
the PyCrypto operation (as we did for all other public key operations).
}
\mypar{Running time of cryptographic operations in total protocol time}
To understand the running times of cryptographic operations among total protocol time,
we measure these times for each protocol in Table~\ref{tab-examp}, 
%the running time of the protocol, and of all calls to cryptographic functions.  
and we show the contributing factors by counting the number of calls to different cryptographic functions and the number of messages passed.

For protocol time, we measure the time used by each role excluding
process setup time, and sum over all roles. 
For the time of all cryptographic operations, which we call library time, 
we measure the time of each \projname{} function call and sum
over all calls.  We collect the counts of calls and messages for the measured execution of each protocol.

Table~\ref{tab-time3} shows the results, grouped by the kinds of cryptographics functions called and sorted by decreasing library time in each group.
Cryptogrpahic functions are listed in the order of expensive ones first: 
modular exponentiation (\co{pow}) for Diffie-Hellman, RSA functions, 
elliptic curve (EC) functions, and shared key functions (SK); 
among RSA functions, \co{keygen}, \co{decrypt} and \co{sign} that use private keys, 
and \co{encrypt} and \co{verify} that use public keys; among EC functions,
\co{keygen} and the rest.
%
%We also count the number of calls to cryptographic functions
%that occur during protocol execution---(1) calls to \co{pow} to do modular
%exponentiation for Diffie-Hellman; (2) calls to RSA functions with separate
%counts for \co{keygen}, \co{decrypt} and \co{sign} that use private keys, and
%\co{encrypt} and \co{verify} that use public keys; (3) calls to elliptic
%curve functions with separate counts for \co{keygen} and others; (4) calls to shared key
%functions---as well as the number of messages passed during protocol execution.
%
%CMK: Add line explaining why SDH running time is so large; pow with large inputs.
\newcommand{\capTabTimeIII}{Number of calls to diferent cryptographic functions
(with expensive ones first),
%\co{pow}, RSA \co{keygen}, private key-using and public key-using 
%functions, and Elliptic Curve (EC) \co{keygen} and other functions. 
CPU times (in milliseconds) of protocol run and library calls, difference between the two times, and number of messages passed.
An empty entry denotes 0.
$^*$ for TLS1.2 indicates that the number of messages can differ when 
different branching conditions hold; our experiments used the condition for 
which 9 messages are passed.
%, which is still more than other protocols and 
%causes the largest time difference.
}
%Percentage of CPU time (in milliseconds)
%ANNIE: seconds? CMK: Yes.
%ANNIE: if so, make ms. and keep precision to 2 digits after.  they got to at least all line up too.
%CMK: Done
%in cryptographic abstractions library (for DHKE-1 and SDH we include an expensive call to the Python pow function)
%ANNIE: (including the power function for DHKE-1 and SDH)
%CMK: Done
%averaged over 20 runs of each protocol.
%annie: rewrite to be simpler and clearer as in Table 7
\begin{table*}[!t]
\forSP{\renewcommand{\arraystretch}{1.3}}
\forSP{\caption{\capTabTimeIII}}
\centering
%\resizebox{\textwidth}{!}{
\begin{tabular}{@{~}l||r|r@{~~}r@{~~}r|r@{~~}r|r||r||r||r||r@{~}}
\multirow{3}{*}{Protocol} & \multirow{3}{*}{\co{pow}} &
\multicolumn{3}{c|}{RSA} & \multicolumn{2}{c|}{EC} & \multirow{3}{*}{SK} &
\multirowcell{3}{Library\\ time} & \multirowcell{3}{Protocol\\ time} & 
\multirowcell{3}{Time\\ diff.} & \multirow{3}{*}{Messages} \\
\cline{3-7}
 & 
 & \multirowcell{2}{key-\vspace{-.5ex}\\gen} 
 & \multirowcell{2}{dec.\vspace{-.5ex}\\sign} 
 & \multirowcell{2}{enc.\vspace{-.5ex}\\veri.} 
 & \multirowcell{2}{key-\vspace{-.5ex}\\gen} & \multirow{2}{*}{rest} & & & &\\
 &&&&&&&&&&&\\
\hline\hline
SDH      & 2 & 2 & 2 & 2 &    &    &    & 160.08 & 161.71 &  1.63  & 3  \\%
DHKE-1   & 2 & 2 & 2 & 4 &    &    &    &  42.60 &  45.89 &  3.29  & 3  \\%
\hline        
NS-PK    &   &   & 5 & 5 &    &    &    &  19.87 &  27.87 &  7.99  & 7  \\%
DS       &   &   & 4 & 6 &    &    &  1 &  17.59 &  22.76 &  5.17  & 3  \\%
DS Simp  &   &   & 2 & 2 &    &    &  3 &   8.34 &  10.40 &  2.06  & 2  \\%
TLS 1.2  &   &   & 1 & 3 &    &    & 33 &   6.90 &  20.33 & 13.43  & 9-13* \\%
\hline
Signal   &   &   &   &   & 11 & 19 & 29 &   8.40 &  19.05 & 10.65  & 8  \\%
DR       &   &   &   &   &  6 &  9 & 25 &   4.93 &   9.70 &  4.77  & 6  \\%
X3DH     &   &   &   &   &  5 & 10 &  4 &   3.72 &   8.42 &  4.70  & 4  \\%
\hline
KRBv5    &   &   &   &   &    &    & 26 &   0.75 &   8.45 &  7.70  & 6  \\%
NS-SK    &   &   &   &   &    &    & 11 &   0.61 &  11.61 & 10.99  & 7  \\%
\hline
\end{tabular}
%}%end \resizebox
\forArxiv{\caption{\capTabTimeIII}}
\forPrelim{\caption{\capTabTimeIII}}
\forACM{\caption{\capTabTimeIII}\vspacebelowfloat}
\label{tab-time3}
\end{table*}
\notes{
%CMK: 10/17/2018 - This table was replaced with one that added a column for
%CMK: shared key encryption, moved the messages column further toward the back,
%CMK: and reversed the order of the protocol time and library time columns.
\begin{table*}[!t]
\forSP{\renewcommand{\arraystretch}{1.3}}
\forSP{\caption{\capTabTimeIII}}
\centering
\resizebox{\textwidth}{!}{
\begin{tabular}{@{}l||r||r|rrr|rr||r|r|r@{}}
\multirow{2}{*}{Protocol} & \multirow{2}{*}{Messages} & 
\multirow{2}{*}{\co{pow}} & \multicolumn{3}{c|}{RSA} &
\multicolumn{2}{c|}{EC} & \multirow{2}{*}{Protocol time} & 
\multirow{2}{*}{Library time} & \multirow{2}{*}{Time diff.} \\
% & Percentage \\
\cline{4-8}
 & & & keygen & dec/sign & enc/veri & keygen & other & & & \\
\hline\hline
SDH      & 3     & 2 & 2 & 2 & 2 &    &    & 161.71 & 160.08 &  1.63  \\%
DHKE-1   & 3     & 2 & 2 & 2 & 4 &    &    &  45.89 &  42.60 &  3.29  \\%
\hline        
NS-PK    & 7     &   &   & 5 & 5 &    &    &  27.87 &  19.87 &  7.99  \\%
DS       & 3     &   &   & 4 & 6 &    &    &  22.76 &  17.59 &  5.17  \\%
DS Simp  & 2     &   &   & 2 & 2 &    &    &  10.40 &   8.34 &  2.06  \\%
TLS 1.2  & 9-13* &   &   & 1 & 3 &    &    &  20.33 &   6.90 & 13.43* \\%
\hline
Signal   & 8     &   &   &   &   & 11 & 19 &  19.05 &   8.40 & 10.65  \\%
DR       & 6     &   &   &   &   &  6 &  9 &   9.70 &   4.93 &  4.77  \\%
X3DH     & 4     &   &   &   &   &  5 & 10 &   8.42 &   3.72 &  4.70  \\%
\hline
KRBv5    & 6     &   &   &   &   &    &    &   8.45 &   0.75 &  7.70  \\%
NS-SK    & 7     &   &   &   &   &    &    &  11.61 &   0.61 & 10.99  \\%
\hline
\end{tabular}
}%end \resizebox
\forArxiv{\caption{\capTabTimeIII}}
\forPrelim{\caption{\capTabTimeIII}}
\label{tab-time3}
\end{table*}
}%end \notes
\notes{
%CMK: 10/16/2018 - replaced this table with one that includes separate column 
%for keygen
\begin{table*}[!t]
    \forSP{\renewcommand{\arraystretch}{1.3}}
    \forSP{\caption{\capTabTimeIII}}
    \centering
    %\resizebox{\columnwidth}{!}{
    \begin{tabular}{@{}l||r@{~~~}||r|r|r||r|r|r@{}}
        Protocol   & Messages & \co{pow} & RSA & EC & Protocol time & Library time & Time diff. \\% & %Percentage \\
        \hline\hline
        SDH        & 3 & 2 & 6 & 0 & 161.71    & 160.08   & 1.63  \\%     & 
        %98.99 \\
        DHKE-1     & 3 & 2 & 8 & 0 & 45.89     & 42.60    & 3.29  \\%     & 
        \hline
        %92.83 \\
        NS-PK      & 7 & 0 & 10 & 0 & 27.87     & 19.87    & 7.99  \\%    & 
        %71.32  \\
        DS         & 3 & 0 & 10 & 0 & 22.76     & 17.59    & 5.17  \\%     & 
        %77.28 \\
        TLS 1.2    & 9-13* & 0 & 4 & 0 & 20.33   & 6.90     & 13.43* \\%    & 
        %33.95  \\
        DS Simp    & 2 & 0 & 4 & 0 & 10.40     & 8.34     & 2.06  \\%     & 
        %80.16 \\
        \hline
        Signal     & 8 & 0 & 0 & 30 & 19.05     & 8.40     & 10.65 \\%    & 
        %44.11  \\
        DR         & 6 & 0 & 0 & 15 & 9.70      & 4.93     & 4.77  \\%     & 
        %50.86 \\
        X3DH       & 4 & 0 & 0 & 15 & 8.42      & 3.72     & 4.70  \\%     & 
        %44.17 \\
        \hline
        NS-SK      & 7 & 0 & 0 & 0 & 11.61     & 0.61     & 10.99 \\%    & 
        %5.29   \\
        KRBv5      & 6 & 0 & 0 & 0 & 8.45      & 0.75     & 7.70  \\%    & 
        %8.84   \\
        \hline
    \end{tabular}
    %}
    \forArxiv{\caption{\capTabTimeIII}}
    \forPrelim{\caption{\capTabTimeIII}}
    \label{tab-time3}
\end{table*}
}%end \notes
\notes{
%CMK: 09/01/2018 - Old Table - Before we added details about public key
%CMK: crypto use.
\newcommand{\capTabTimeIII}{Number of messages passed and CPU times (in 
milliseconds)
    for running a protocol and all \projname{} library functions executed in 
    that run,
    and the difference between protocol time and library time. Longer running 
    times of SDH
    and DHKE-1 are due to calls to expensive \co{pow} function with 
    large inputs.}
%\begin{table}[htb]
\begin{table}[!t]
\forSP{\renewcommand{\arraystretch}{1.3}}
\forSP{\caption{\capTabTimeIII}}
\centering
%\resizebox{\columnwidth}{!}{
\begin{tabular}{@{}l|r||r|r||r@{}}
Protocol   & \# msgs & Protocol time & Library time & Diff. \\% & Percentage \\
\hline\hline
SDH        & 3  & 161.71    & 160.08   & 1.63  \\%     & 98.99 \\
DS Simp    & 2  & 10.40     & 8.34     & 2.06  \\%     & 80.16 \\
DHKE-1     & 3  & 45.89     & 42.60    & 3.29  \\%     & 92.83 \\
\hline
X3DH       & 4  & 8.42      & 3.72     & 4.70  \\%     & 44.17 \\
DR         & 6  & 9.70      & 4.93     & 4.77  \\%     & 50.86 \\
DS         & 3  & 22.76     & 17.59    & 5.17  \\%     & 77.28 \\
\hline
KRBv5      & 6  & 8.45      & 0.75     & 7.70  \\%    & 8.84   \\
NS-PK      & 7  & 27.87     & 19.87    & 7.99  \\%    & 71.32  \\
Signal     & 8  & 19.05     & 8.40     & 10.65 \\%    & 44.11  \\
NS-SK      & 7  & 11.61     & 0.61     & 10.99 \\%    & 5.29   \\
TLS 1.2    & 9-13 & 20.33   & 6.90     & 13.43 \\%    & 33.95  \\
\hline
\end{tabular}
%}
\forArxiv{\caption{\capTabTimeIII}}
\label{tab-time3}
\end{table}
}
\notes{
%CMK: 08/20/2018 - Old table data.
\begin{table}[htbp]
\centering
\resizebox{55ex}{!}{
\begin{tabular}{ l|r||r|r||r }
Protocol   & \# messages & Protocol time & Library time & Difference \\
% & Percentage \\
\hline\hline
DS Simp    & 2  & 10.42     & 8.38    & 2.04 \\%     & 80.42    \\
SDH        & 3  & 161.77    & 159.35   & 2.42 \\%     & 98.5    \\
DHKE-1     & 3  & 45.99     & 42.68    & 3.31 \\%     & 92.8    \\
\hline
X3DH       & 4  & 8.38      & 3.69     & 4.69 \\%     & 44.03    \\
DR         & 6  & 9.77      & 4.86     & 4.91 \\%     & 49.74    \\
DS         & 3  & 23.24     & 17.29    & 5.95 \\%     & 74.4    \\
\hline
KRBv5      & 6  & 8.47      & 0.74     & 7.73 \\%      & 8.74     \\
NS-PK      & 7  & 27.84     & 19.80     & 8.04 \\%     & 71.12    \\
Signal     & 8   & 19.22    & 8.45     & 10.77 \\%    & 43.96    \\
NS-SK      & 7  & 11.59     & 0.62     & 10.97 \\%     & 5.35     \\
TLS 1.2    & 9-13 & 20.49   & 6.90      & 13.59 \\%    & 33.67   \\
\hline
\end{tabular}}
\caption{Number of messages involved and CPU times (in milliseconds) for running each protocol and \projname{} library functions
during that run, and the difference between protocol time and library time.}
%Percentage of CPU time (in milliseconds)
%ANNIE: seconds? CMK: Yes.
%ANNIE: if so, make ms. and keep precision to 2 digits after.  they got to at least all line up too.
%CMK: Done
%in cryptographic abstractions library (for DHKE-1 and SDH we include an expensive call to the Python pow function)
%ANNIE: (including the power function for DHKE-1 and SDH)
%CMK: Done
%averaged over 20 runs of each protocol.
%annie: rewrite to be simpler and clearer as in Table 7
\label{tab-time3}
\end{table}
}

For library time, we see that it is almost fully determined by the counts of calls to more expensive functions, with two exceptions: 
(1) SDH and DHKE-1 both have the same numbers
of expensive calls, especially power function \co{pow} to compute Diffie-Hellman shared secrets, but the larger time for SDH is because it uses values that are 3 times as large; 
(2) Signal uses EC, but it has many more calls 
to EC \co{keygen} and thus a slightly larger library time than DS Simp and TLS 1.2 that use RSA but have few calls of non-\co{keygen} functions.
In fact, with the exception of Signal, 
the library time is sorted completely in decreasing order.

%Two main factors contribute to total protocol time: use of public key 
%cryptography and number of messages passed.
%
%For running time of \projname{}
%functions, the main contributing factor, aside from the bare number of
%cryptographic operations, is how many of them are public key operations.
%Together the number of protocol messages and the number of public key 
%operations suffice to explain the results in Table~\ref{tab-time3}.

Protocol time is also mostly in decreasing order, but with 
three exceptions: TLS 1.2, Signal, and NS-SK.  
This is because protocol time is also affected by the number of messages passed during the protocol run.  In fact, the three exceptions are from protocols that have the most messages.  

We consider the difference between protocol and library times. We see that for each group, a larger time difference corresponds to a larger number of messages, with one exception: SDH and DHKE-1 both have 3 messages, but 
the larger difference for DHKE-1 is due to additional local, non-cryptographic
computations in DHKE-1 but not in SDH.

\notes{
Use of RSA public key functions are the next most expensive among library 
operations, and this explains the second group of protocols; the first two are 
more expensive because they have more function calls. Though 
NS-PK and DS use 
the same number of calls to public key functions, NS-PK features one more call 
to an expensive private key-using operation (\co{decrypt} or \co{sign}) than 
DS, which accounts for the difference in their library times. The same 
explanation applies to the difference in library time between TLS 1.2 and DS 
Simp.

Elliptic curve functions used by X3DH, DR, and Signal in the third group are 
more efficient than RSA functions, which is why their cryptographic operation 
times are smaller. Though DR and X3DH make the same number of 
calls to EC 
functions, DR uses one more expensive key generation operation, which explains 
the larger library time of DR.

The last group uses no public key functions and has minimum library time.

The number of messages passed during protocol execution also contributes 
directly to the total protocol time, but it is not as significant as calls to 
expensive functions such as \co{pow}.  We see this effect within each group in 
Table~\ref{tab-time3}.
%However, it can be overridden by greater use of public key cryptographic 
%functions, as can be seen by the position of SDH and DHKE-1 at the top, or 
%the relatively low position of TLS in the RSA only group.
}

\notes{
Protocols with 3 or fewer messages like simplified DS, SDH, and DHKE-1 have
small differences between total time and \projname{} functions time whereas
protocols with 6 or more messages like TLS, Signal, NS-SK, and Kerberos exhibit
larger differences, indicating a dominance of message passing over cryptographic
operation time. For protocols in the middle, a driving factor is the use of
shared vs.\ public key cryptography. For example, while both NS-SK and NS-PK
involve the same number of messages, the difference between total and library
time is higher for NS-SK than for NS-PK.
}

\notes{
%CMK: 09/01/2018 - Added public key crypto call count to table, so this
%CMK: explanatory paragraph can be shortened.
Protocols with 3 or fewer messages like simplified DS, SDH, and DHKE-1 have
small differences between total time and \projname{} functions time whereas
protocols with 6 or more messages like TLS, Signal, NS-SK, and Kerberos exhibit
larger differences, indicating a dominance of message passing over cryptographic
operation time. For protocols in the middle, a driving factor is the use of
shared vs.\ public key cryptography. For example, while both NS-SK and NS-PK
involve the same number of messages, the difference between total and library
time is higher for NS-SK than for NS-PK. Another contributing factor is the
type of public key encryption used. For example, Double Ratchet involves 6
messages and NS-PK inolves 7 but, because elliptic curve cryptography used
by DR is more efficient than the RSA operations used by NS-PK, DR has a smaller
overall protocol time and \projname{} function calls take up a smaller
proportion of the total protocol time.
}%

\notes{
Consider the three protocols with the smallest difference between protocol time and library time: simplified DS, SDH, and
DHKE-1. Each of these protocols features a small number of messages---two for simplified DS, and three each for SDH and DHKE-1.
They also make use of expensive public key operations. As a result, most of their protocol running time is spent executing these
public key operations and relatively little time is spent on message passing. That is why the difference between protocol time and
library time is small for these protocols (for SDH and DHKE-1, their long running times are a result of expensive modular
exponentiation operations used to compute a Diffie-Hellman shared secret).

Protocols that feature a larger number of messages or that make greater use of shared key cryptography should display
a larger difference between protocol running time and library running time. This is what we observe. The third and fourth
largest differences between protocol and library time are measured for NS-SK and Kerberos. Both of these protocols involve
more messages (seven for NS-SK, six for Kerberos) and make exclusive use of shared key cryptography.

We must still account for the protocols in the middle, and for the exceptional case of TLS 1.2 which has the largest
measured difference between protocol time and library time. For middle of the road protocols like NS-PK, the
most plausible explanation is that the number of messages contributes more strongly to protocol time than the use of public
key cryptography. NS-PK involves seven message exchanges, and the time required for message passing dominates the time used
for public key operations, which results in a large gap between protocol time and execution time. For another middle--difference
protocol, X3DH, part of the explanation is the number of messages (four of X3DH). However, an equally important part of the
explanation is that X3DH, like DR and Signal, makes use of elliptic curve cryptography for its public key operations, and
elliptic curve operations are more efficient than the RSA cryptography used by the other public key protocols.

Finally, in the case of TLS 1.2, we can again appeal to the number of messages. As implemented, our TLS 1.2 involves nine messages
(and can use up to thirteen, if all optional messages are included). This is more than any of the other protocols. In addition, 
our implementation makes limited use of public key cryptography---several calls to verify to check signatures on certificates and
encryption to protect the pre-master secret sent by the client.  All other cryptographic operations in our implementation 
are shared key operations.
}

\notes{
\begin{table}[htbp]
\centering
\begin{tabular}{ l||r|r||r }
Protocol   & Protocol time & Library time & Difference \\% & Percentage \\
\hline\hline
NS-SK      & 10.4010       & 0.6082       & 9.7928 \\%     & 5.85     \\
\hline
NS-PK      & 27.3277       & 20.0818      & 7.2459 \\%     & 73.49    \\
\hline
DS Simp    & 10.4335       & 8.3694       & 1.7941 \\%     & 80.22    \\
\hline
DS         & 23.0185       & 17.3968      & 5.6217 \\%     & 75.58    \\
\hline
%OR         & 6.2149        & 0.5392       & 5.6757 \\%     & 8.68     \\
%\hline
%WL         & 10.6516       & 0.6515       & 10.0001 \\%    & 6.12     \\
%\hline
%Ya         & 5.2508        & 0.4896       & 4.7612 \\%     & 9.32     \\
%\hline
DHKE-1     & 46.1504       & 42.6002       & 3.5502 \\%     & 92.31    \\
\hline
SDH        & 162.0064      & 159.9504      & 2.0560  \\%     & 98.73    \\
\hline
X3DH       & 8.3644        & 3.7777        & 4.5867 \\%     & 45.16    \\
\hline
DR         & 9.7677        & 4.9338        & 4.8339 \\%     & 50.51    \\
\hline
Signal     & 19.0312       & 8.4818        & 10.5494 \\%    & 44.57    \\
\hline
KRBv5      & 8.3483        & 0.7393        & 7.6090 \\%      & 8.86     \\
\hline
TLS 1.2    & 20.6099       & 6.8720        & 13.7379 \\%    & 33.34    \\
\hline
\end{tabular}
\caption{CPU times (in milliseconds) for running each protocol and \projname{} library functions
during that run, and the difference between protocol time and library time.}
%Percentage of CPU time (in milliseconds)
%ANNIE: seconds? CMK: Yes.
%ANNIE: if so, make ms. and keep precision to 2 digits after.  they got to at least all line up too.
%CMK: Done
%in cryptographic abstractions library (for DHKE-1 and SDH we include an expensive call to the Python pow function)
%ANNIE: (including the power function for DHKE-1 and SDH)
%CMK: Done
%averaged over 20 runs of each protocol.
%annie: rewrite to be simpler and clearer as in Table 7
\label{tab-time3}
\end{table}
}

\notes{
%CMK: 02/16/2018 - This is the May, 16 2017 version of this table
\begin{table}[h]
\resizebox{55ex}{!}{
\begin{tabular}{ c||c|c|c }
Protocol   & Protocol Time & Library Time  & Ratio  \\
\hline\hline
Simple DS  & 17.32         & 15.49         & 0.89 \\
%Simple DS  & 0.01731712765 & 0.0154944921  & 0.89 \\
\hline
Full DS & 26.81 & 23.69 & 0.88 \\
%Full DS-PK & 0.02680816055 & 0.023699051   & 0.88 \\
\hline
NS-SK* & 13.48 & 8.11 & 0.60 \\
%NS-SK*     & 0.01348293965 & 0.00811057995 & 0.60 \\
\hline
NS-PK* & 35.03 & 23.73 & 0.68 \\
%NS-PK*     & 0.035034114   & 0.02372638455 & 0.68 \\
\hline
OR & 14.77 & 9.07 & 0.61 \\
%OR         & 0.01476915905 & 0.0090711846  & 0.61 \\
\hline
WL & 18.43 & 10.45 & 0.57 \\
%WL         & 0.0184349853  & 0.0104478436  & 0.57 \\
\hline
Ya & 14.93 & 9.69 & 0.65 \\
%Ya         & 0.01492959955 & 0.0096858349  & 0.65  \\
\hline
DHKE-1 & 51.28 & 28.44 & 0.55 \\
%DHKE-1     & 0.0512816578  & 0.0280808473  & 0.55 \\
\hline
SDH & 170.89 & 155.03 & 0.92 \\
%SDH        & 0.170887113   & 0.0270325829  & 0.16 \\
\hline
KRBv5 & 18.40 & 13.33 & 0.72 \\
%KRBv5      & 0.0183956573  & 0.0133255396  & 0.72 \\
\hline
TLS 1.2 & 44.33 & 29.86 & 0.67 \\
%TLS 1.2    & 0.04432878145 & 0.02986350885 & 0.67 \\
\hline
\end{tabular}}
\caption{Proportion of CPU time (in milliseconds)
%ANNIE: seconds? CMK: Yes.
%ANNIE: if so, make ms. and keep precision to 2 digits after.  they got to at least all line up too.
%CMK: Done
spent in cryptographic abstractions library (for DHKE-1 and SDH we include an expensive call to the Python pow function)
%ANNIE: (including the power function for DHKE-1 and SDH)
%CMK: Done
averaged 
         over 20 runs of each protocol.}
\label{tab-time3}
\end{table}
}

%CMK: 05/08/2017 - Because of the change to the table (tab-time3), where we replaced the percentage difference between
%CMK: protocol time and library time with the real time difference between them, the analysis section
%CMK: needed to be rewritten.
\notes{
Table~\ref{tab-time3} reveals an interesting result. Whether or not the 
cryptographic operations in the \projname{} library account for protocol
overhead depends upon the kind of cryptographic operation in use. For
protocols that make use solely of shared encryption (NS-SK, OR, WL, Ya, and KRBv5),
only 5-10\% of the running time is spent executing \projname{} library functions, and it
is message passing that accounts for the vast majority of the running time of the protocol.

By contrast, for those protocols that rely on public key cryptography (Simple DS, Full DS-PK,
NS-PK, DHKE-1, and SDH), 73-99\% of the protocol time is spent executing those expensive public
key operations (for DHKE-1 and SDH we also include expensive calls to Python's \co{pow} function
in the library time measurement because those calls are used for session key establishment). However,
the remaining protocol time is exactly what one would expect given the number of messages passed between
the roles in protocol. 

TLS stands out with 33\% of its running time spent executing library functions. This is not surprising for two reasons.
(1) Though the basic TLS uses public key methods, it makes limited use of them; encryption and decryption for
the pre-master secret, and verification for certificates. Bulk encryption and integrity protection in the record layer
are done with shared key algorithms. (2) The basic TLS handshake involves nine message exchanges. Most of the public key
protocols involve just two or three message exchanges. Together these two reasons explain why the proportion of time
spent in the \projname{} library falls in between that of the other protocols.
}

\mypar{Comparison with using Python and PyCrypto on NK-SK}
Figure~\ref{fig-nsplot} shows the running times of NS-SK written using
\projname{} and PyCrypto on top of DistAlgo and Python, for all 4 combinations,
measured by repeating NS-SK on increasing numbers of runs.

%\begin{figure}[htbp]
\begin{figure}[!t]
\centering
\hspace{-2ex}\begin{tikzpicture}[scale=\forACM{1.2}\forArxiv{1.1}]
\begin{axis}[
    xlabel={\# protocol runs (in thousands)},
    ylabel={Time (sec)},
    xmin=2, xmax=12,
    ymin=0, ymax=100,
    xtick={2,4,6,8,10,12},
    legend style={font=\small, at={(0,1)}, anchor=north west},
    ymajorgrids=true,
    grid style=dashed,
    nodes={scale=0.85, transform shape},
]

\addplot[
    color=black,
    mark=square,
    ]
    coordinates {(2,15.6626)(4,31.1166)(6,46.4592)(8,61.97)(10,77.2016)(12,92.8744)};
    \addlegendentry{\projname{}+DistAlgo}

\addplot[
    color=black,    
    mark=diamond,
    ]
    coordinates {(2,15.8366)(4,31.1742)(6,47.0072)(8,62.3803)(10,78.9811)(12,94.6295)};
    \addlegendentry{PyCrypto+DistAlgo}

\addplot[
    color=black,
    mark=o,
    ]
    coordinates {(2,3.7911)(4,7.2223)(6,10.3685)(8,14.1523)(10,17.1945)(12,18.8727)};
   \addlegendentry{\projname{}+Python}

\addplot[
    color=black,
    mark=triangle,
    ]
    coordinates {(2,3.3668)(4,6.5604)(6,9.5333)(8,12.205)(10,14.7972)(12,17.7029)};
   \addlegendentry{PyCrypto+Python}
   
\notes{
%CMK: 08/21/2018 - Old measurements replaced on this date.
\addplot[
    color=black,
    mark=square,
    ]
    coordinates {(2,15.71)(4,31.26)(6,46.83)(8,62.18)(10,77.31)(12,92.84)};
    \addlegendentry{\projname{}+DistAlgo}
    
\addplot[
    color=black,    
    mark=diamond,
    ]
    coordinates {(2,16.03)(4,31.42)(6,46.96)(8,62.09)(10,78.13)(12,92.74)};
    \addlegendentry{PyCrypto+DistAlgo}

\addplot[
    color=black,
    mark=o,
    ]
    coordinates {(2,3.6424)(4,6.9147)(6,9.3826)(8,12.018)(10,15.4148)(12,16.6822)};
   \addlegendentry{\projname{}+Python}

\addplot[
    color=black,
    mark=triangle,
    ]
    coordinates {(2,3.2704)(4,6.1659)(6,8.119)(8,11.7136)(10,13.542)(12,15.5815)};
   \addlegendentry{PyCrypto+Python}
}%end notes
\notes{ 
%CMK: 05/09/2018: Old, superlinear measurement for DA+SA and DA+PC
\addplot[
    color=black,
    mark=square,
    ]
    coordinates {(2,23.34)(4,67.64)(6,130.99)(8,214.64)(10,317.42)};
    \addlegendentry{DistAlgo+\projname{}}

\addplot[
    color=black,    
    mark=,
    ]
    coordinates {(2,22.79)(4,65.49)(6,130.31)(8,210.81)(10,311.28)};
    \addlegendentry{DistAlgo+PyCrypto}
}%end notes

\end{axis}
\end{tikzpicture}
\caption{Running times of NS-SK on increasing numbers of protocol runs.}\forACM{\vspacebelowfloat}
\label{fig-nsplot}
\end{figure}
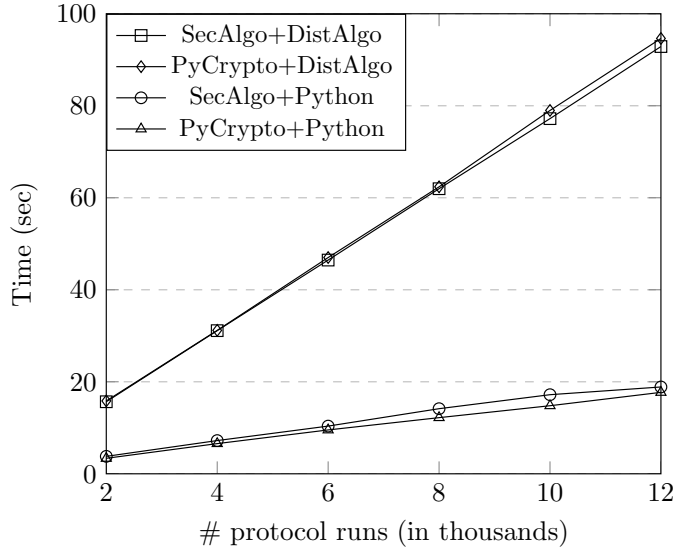

\notes{%annie: commented out 8/21/18
\begin{figure}
\centering
\begin{tikzpicture}
\begin{axis}[
    xlabel={\# protocol runs $/ 10^{3}$},
    ylabel={Time (sec)},
    xmin=2, xmax=12,
    ymin=0, ymax=100,
    xtick={2,4,6,8,10,12},
    legend style={at={(0,1)}, anchor=north west},
    ymajorgrids=true,
    grid style=dashed,
]

%\notes{
%CMK: 08/21/2018 - Old measurements replaced on this date.
\addplot[
    color=black,
    mark=square,
    ]
    coordinates {(2,15.71)(4,31.26)(6,46.83)(8,62.18)(10,77.31)(12,92.84)};
    \addlegendentry{\projname{}+DistAlgo}
    
\addplot[
    color=black,    
    mark=diamond,
    ]
    coordinates {(2,16.03)(4,31.42)(6,46.96)(8,62.09)(10,78.13)(12,92.74)};
    \addlegendentry{PyCrypto+DistAlgo}

\addplot[
    color=black,
    mark=o,
    ]
    coordinates {(2,3.6424)(4,6.9147)(6,9.3826)(8,12.018)(10,15.4148)(12,16.6822)};
   \addlegendentry{\projname{}+Python}

\addplot[
    color=black,
    mark=triangle,
    ]
    coordinates {(2,3.2704)(4,6.1659)(6,8.119)(8,11.7136)(10,13.542)(12,15.5815)};
   \addlegendentry{PyCrypto+Python}
%}%end notes
\notes{ 
%CMK: 05/09/2018: Old, superlinear measurement for DA+SA and DA+PC
\addplot[
    color=black,
    mark=square,
    ]
    coordinates {(2,23.34)(4,67.64)(6,130.99)(8,214.64)(10,317.42)};
    \addlegendentry{DistAlgo+\projname{}}

\addplot[
    color=black,    
    mark=,
    ]
    coordinates {(2,22.79)(4,65.49)(6,130.31)(8,210.81)(10,311.28)};
    \addlegendentry{DistAlgo+PyCrypto}
}%end notes

\end{axis}
\end{tikzpicture}
\caption{(OLD)Running times of NS-SK on increasing numbers of protocol runs.}
\label{fig-nsplot}
\end{figure}
}

All 4 implementations show a linear increase in running time as the number of
runs increases. The difference between using \projname{} and using PyCrypto, on top of
DistAlgo or on top of Python, is small: at most 2.4 seconds and between -2\% and 16\%.
The difference can sometimes be negative because of the small overhead of SecAlgo and the usual variation in running times of multi-process protocols even when averaged over 50 runs.
%The small negative difference often happens for using DistAlgo; we have not been able to identify a particular reason.
%Same holds for using \projname{} and PyCrypto on top of Python.

Using DistAlgo is about 5 times as slow as Python, but that is expected and is
the subject of DistAlgo compilation and optimizations studied 
separately~\cite{liu2017clarity}.%
%Annie: tt wasn't 5-6 but maybe 2-3 then:
%CMK: I know that the constant factor difference reported in the OOPSLA paper was smaller. I've been wondering
%CMK: why we get a larger difference here. Mostly I wanted the citation to support the claim that this overhead
%CMK: could be eliminated through optimization.
%A similar constant factor difference was measured by the authors
%of DistAlgo, who assure us this difference can be eliminated by optimization \cite{liu2017clarity}.

The main result is that whether using DistAlgo or Python, using \projname{} is
at most a small increase over using PyCrypto directly, while being
safe and much simpler %easier
to use.%

% scott: removed for PLAS 2019 submission; as an OOPSLA 2019 reviewer mentions, this is primarily about the programming languages, not the crypto libraries.
\notes{We also measured the running times of NS-SK in C\# and Java. The C\# program is
about twice as fast as the PyCrypto+Python implementation, while the Java program
is about twice as slow as the PyCrypto+Python implementation. Both are
faster than the \projname{}+DistAlgo implementation.
We did not study them further partly due to the much more complex and
unattractive code compared to DistAlgo and Python, but also because
the performance of cryptographic libraries in Python is becoming more competitive
by library code being in C.}

\notes{
%CMK: 05/10/2018 - This was the description for the original plot with the superlinear results for the protocols
%CMK: built on top of DistAlgo.
The lines for DA+SA and DA+PC in Figure~\ref{fig-nsplot} nearly overlap. The gap between the two varies between
0.54 seconds at 2000 rounds and 6.15 seconds at 6000 rounds, and appears to grow very slowly as the number of
rounds are increased. Most of the growth in the running time of DA+SA and DA+PC can be attributed to the
increasing burden on DistAlgo. In particular, DistAlgo maintains message histories for each process. Since the
repetition of the protocol is done without terminating the processes involved, the message history grows larger
and larger as the number of rounds increases. Any query over this message history--a common operation when
using DistAlgo's \co{await} abstraction would be impacted by the growth in the message history.

By contrast, the Py+PC curve grows very slowly. At 2000 rounds, the total running time of DA+SA and DA+PC are
both about seven times greater than that of Py+PC. By 10000 rounds, the DA+SA and DA+PC are running two hundred
times slower than Py+PC. Again, we think that this difference can be mostly explained by the burden additional
rounds place on DistAlgo's message handling mechanisms. The Python implementation uses simple UDP sockets for
communication and does not record any message history, which is why the increase in rounds, which causes the
increase in the number of messages, has much less impact on the running time of Py+PC.
}

\notes{
%CMK: 02/14/2018 - This table and its accompanying description will be replaced
%CMK: with a figure containing curves that describe CPU Time as the size of the
%CMK: session key is increased.
%CMK: 05/07/2018 - Actually, we are replacing this table and its discussion with
%CMK: a plot of the running time (CPU time) of DA+SA, DA+PC, and Py+PC versions
%CMK: of NS-SK. We vary the number of rounds the protocol runs, and then plot
%CMK: the total CPU time required to run the protocol that many times (rather
%CMK: than taking the average in order to get an accurate estimate of the time
%CMK: used to execute the protocol once).
For each protocol implementation, we measure the running time of both
the protocol and time spent executing cryptographic library code.
The results are given in Table~\ref{tab-time2}.

\begin{table}[h]
\centering
\resizebox{55ex}{!}{
\begin{tabular}{l|l|r|r}
Language & Crypto library           & Protocol Time & Library Time \\
\hline
C        & Openssl \cite{OpenSSL}   & ???           &              \\
\hline
C\#      & .NET Cryptography        & ???           &              \\
\hline
Java     & JCA \cite{JCA}           & ???           &              \\
\hline
Python   & PyCrypto \cite{PyCrypto} & 1.3962        &  0.5271      \\
\hline
DistAlgo & PyCrypto                 & 10.1953       &  0.5814      \\
\hline
DistAlgo & \projname{}              & 10.4010       &  0.6082      \\
\end{tabular}}
\caption{Average CPU time to execute alternative
         implementations of the NS-SK protocol.}
\label{tab-time2}
\end{table}

As expected, the C implementation is the most efficient running \% faster than the \projname{} implementation.
Both the Java, \% faster than \projname{}, and C\#, \% faster than \projname{} implementations run more quickly
than the Python implementation, which we expect because both of those languages are compiled to bytecode before
execution, whereas Python is an interpreted language. The \projname{} plus DistAlgo program, which 
compiles to Python, is 7.44 times slower than the manually written Python implementation. The use of \projname{}
abstractions accounts for only 0.9\% of the overhead. The bulk of the difference in efficiency between the \projname{} plus DistAlgo
and Python implementations is due to the much simpler architecture of the Python implementation (plain multiprocessing
and UDP message passing, without any threading) when compared to DistAlgo. The DistAlgo authors report that DistAlgo programs
will be at least 3 times as slow as manually-written Python programs~\cite{Liu:2012:CED:2398857.2384645}, but they claim 
that this overhead can be optimized away. Given that Python is widely used, the additional overhead of writing the
implementation in DistAlgo using \projname{} is reasonable.
}

\section{Related work and conclusion}
\label{sec-related}
%annie-open: 12/17/2017
%annie: pls remember to use/relate to and cite the paper you found on Comparing the Usability of Cryptographic APIs

\notes{
\begin{itemize}
\item
  Security protocol specification languages, since that is what we are
  comparing ourselves to.

\item
  Improved libraries, as listed and described in my RPE. Since they make the
  same sort of contribution I am making.

\item
  Anything else?
\end{itemize}
}
There have been many efforts at building better cryptographic libraries
providing simpler interfaces. These include the
NaCl library~\cite{NaClSource, Bernstein:2012:SIN:2406723.2406735} for C and
C++; libsodium \cite{libsodium-Source}, a portable version of NaCl with a
slightly improved interface; the pyca/cryptography
library~\cite{cryptography-ioSource} for Python; the Charm
library~\cite{CharmSource, charm13} for Python; the Keyczar
library~\cite{KeyczarSource, DeyWeis2015} for C++, Java, and Python; and the
Tink library~\cite{Tink} for C++, Java, Go, and Objective-C.

These libraries simplify use of cryptographic operations in ways similar to
\projname{}. Simplifying techniques to reduce the number of decisions for users 
include: (1) requiring fewer inputs from the user, (2) handling tedious, 
routine tasks automatically behind the scenes, (3) supporting better default 
configurations, and (4) removing unsafe
algorithms and implementations.

However, these libraries fall short when compared with \projname{}. Charm 
provides
simplified use for only a single operation---shared key authenticated 
encryption.
pyca/cryptography provides simplified use of only shared key authenticated
encryption and X.509 certificate handling. Keyczar and Tink provide only
shared key authenticated encryption, hybrid encryption, %digital
signing, and
message authentication code creation. NaCl and libsodium provide a much more 
complete
set of cryptographic operations, but provide little to no configurability but
only one algorithm for most cryptographic operations.

%CMK: 04/2018 - I have tried to compress the separate entries given for each
%CMK: library or API listed below, into a couple of paragraphs that describe all
%CMK: of them.
\notes{
The NaCL (pronounced ``salt'') project \cite{NaClSource, Bernstein:2012:SIN:2406723.2406735}
is a recent cryptographic library for C and C++. It offers a simpler interface, a curated
selection of cryptographic algorithms, and more secure and efficient
implementations of selected algorithms. However the NaCl interface offers little support
for configuration or choosing safe, non-default algorithms included in the their library.

The Charm project \cite{CharmSource, charm13} provides a framework for rapidly prototyping new
cryptosystems---cryptographic algorithms or combinations of algorithms. In
addition to a host of mathematical resources for developing new cryptosystems,
they offer a new cryptographic library, with a simpler, higher-level interface.
For -key encryption, they provide a similar abstraction that makes IV
generation and other details transparent to the user.  However, they offer no
new abstraction for asymmetric-key encryption.

pyca/cryptography \cite{cryptography-ioSource} is cryptographic library for Python that
aims to provide a simpler, more usable interface for cryptographic operations.
It is split into two ``layers'', the \textit{recipes} layer and the \textit{hazmat} layer.
The recipes layer contains simple, high-level cryptographic interfaces and
safe defaults for shared key authenticated encryption and X.509 certificate handling.
Unfortunately, it only provides a high-level interface for these two operations.
All other cryptographic operations must be accessed through the low-level, hazmat layer, which uses
the conventional expertise-requiring interface.

The projects closest to ours in intent and result are Keyczar 
\cite{KeyczarSource, DeyWeis2015} and Tink \cite{Tink}, both of which are developed
by researchers at Google.

Keyczar is described as an
open-source cryptographic toolkit for Java, C++, and Python. Keyczar offers a 
simple interface for both shared key and public key cryptography, automated 
handling of IVs and other low-level details, safe default choices for 
algorithms, modes of operation, and key sizes.  All configuration in 
Keyczar---choice of cryptographic algorithm, mode of operation, key size, 
etc.---is done at the time of key generation, but when using Keyczar all keys must
be generated using a standalone tool included in their distribution. This 
significantly complicates configuration as well as dynamic generation of keys,
which is a feature of all key exchange security protocols.

The Tink cryptographic library is available for Java, C++, Go, and Objective-C
(the Java implementation is being used in production, the others are still in
development). It is similar to Keyczar in several ways. It provides a simple
interface for accessing cryptographic operations through the instantiation of
cipher objects upon which one calls cryptographic methods. Tink also performs
misuse-mitigation by hiding low-level details behind its interface. Finally,
Tink handles configuration through key generation. There are two significant
differences between Tink and Keyczar. First, Tink supports dynamic key
generation through its interface (they also provide a  command line tool for key
generation and mangagement). Second, Tink offers a much more limited selection
of cryptographic operations, which they call ``primitives''---authenticated
encryption with associated data (AEAD), message authentication (MAC), digital
signitures, and hybrid encryption---most of which use a combination of 
cryptographic operations to accomplish a specific security-related task. This
limitation makes it impossible to use Tink to write many of security protocols
we have discussed above.
}

Acar et al.~\cite{AcarAPIUsability2017} study the usability of five Python
cryptographic libraries: PyCrypto~\cite{PyCrypto},
M2Crypto~\cite{M2Crypto-Source}, Keyczar~\cite{KeyczarSource, DeyWeis2015},
pyca/cryptography~\cite{cryptography-ioSource}, and PyNACL~\cite{PyNaCl-Source} (a
Python binding of NACL). 
They found that 
%users had more difficulty
%writing working solutions to their challenges when using simplified APIs.
%Instead, %annie: conflict with what you wrote below
%the presence of 
clear documentation and concrete code examples were the
most significant factors determining whether subjects %users
produced solutions that work.
%However, for working solutions, those 
Furthermore, they found that code written with simplified APIs were much
more likely to be secure, while %the working solutions 
code written with low-level
libraries were more likely to contain mistakes that compromised their security.
%, because simplified APIs provided fewer opportunities for cryptographic misuse.
\projname{} provides higher-level, simpler APIs than these previous libraries. 

Egele et al. \cite[p. 81]{Egele:2013} study cryptographic misuse in Android and propose
mitigation strategies: (1) introduction of better default configurations in
cryptographic libraries and (2) provision of better, more complete documentation
of cryptographic libraries. \projname{} realizes the first by default
configurations that implement best security practice and allows the second to 
be made much simpler and easier to use.  
%As explained in
%Section~\ref{sec-motive}, while better documentation would be a welcome
%improvement, educating developers is not sufficient. 

FixDroid~\cite{Nguyen:2017} is an IDE plug-in for the Android SDK that
identifies cryptographic mistakes in source code, as it is written, and
provides suggested corrections. CogniCrypt~\cite{Kruger:2017} automatically
generates Java code for a collection of common cryptographic tasks (e.g.,
encrypting data with a password, storing passwords, secure communication, etc.) 
and performs static analysis to verify that generated code is properly
integrated into the user's application. CDRep~\cite{Ma:2016} acts directly on
Android binaries by using static analysis to detect cryptographic misuses and
then generates and applies patches to correct them.
Use of \projname{} allows many tasks of such tools to be greatly simplified or completely eliminated.

Security protocol specification languages are for abstract formulation and verification
of security protocols.
\notes{
%CMK: 02/2018 - Removed because unnecessary. The terms symbolic and computational model
%CMK: have citations.
All but one of the specification languages discussed operate in the 
symbolic model for security protocol specification, in which cryptographic
functions are black boxes that operate on contentless terms
\cite{Abadi2002, Blanchet:2012:SPV:2260577.2260579} and the
adversary is constrained to use only the defined cryptographic operations to
manipulate messages.  The exception, Cryptoverif \cite{CryptoVerifSource, BlanchetTDSC07}, operates in the 
computational model, in which messages are bitstrings and cryptographic
operations are functions on bitstrings. In this model, the adversary can be any
polynomial-time Turing machine \cite{BlanchetTDSC07}.
}
Scyther~\cite{ScytherSource, Cr2008Scyther}, AVISPA~\cite{AVISPA_Source,
Armando:2005:ATA:2153230.2153265}, ProVerif \cite{ProVerif_Source,
BlanchetFOSAD14}, and CryptoVerif~\cite{CryptoVerifSource, 4358700} are process or role oriented %approach to specifying protocols 
similar to \projname{}
plus DistAlgo. Tamarin~\cite{TamarinSource, MSCB2013Tamarin} models the state of
the protocol %at a time 
as a multi-set of facts and models protocol actions as rewrite
rules operating on these facts.
\notes{
%CMK: 07/09/2018 - Deemed irrelevant; deleted.
All but one of the specification languages operate in the symbolic
model~\cite{Abadi2002, Blanchet:2012:SPV:2260577.2260579}. The exception,
CryptoVerif operates in the computational model~\cite{BlanchetTDSC07}.
}
%
%All of the languages discussed %in this paper 
%are associated with tools that
%perform automated verification that the specified protocol satisfies certain
%security properties.Scyther specifications are the most concise. CryptoVerif
%specifications can be transformed into a runnable OCAML
%implementations~\cite{6329273}. 
%
%CMK: 04/2018 - Removing individual items for each specification language. It is
%CMK: not necessary to provide this much detail about each one. There relevance
%CMK: to this project is the same in each case--they enable high-level
%CMK: descriptions of protocols, just like the combination of DistAlgo and
%CMK: \projname{}. Their main deficiency is also the same--none of their
%CMK: specifications run.
\notes{
\begin{itemize}

\item \textit{Scyther.} The Scyther \cite{ScytherSource, Cr2008Scyther} specification language and 
verification tool offer role-based specifications of security protocols and
operates in the symbolic model~\cite{Abadi2002, Blanchet:2012:SPV:2260577.2260579}.
The language provides extremely concise, high-level abstractions for 
cryptographic operations and distributed computation, but offers few resources for specifying local
computations and control logic.

\item \textit{AVISPA.} 
The AVISPA \cite{AVISPA_Source, Armando:2005:ATA:2153230.2153265} specification language and verification
tool also offers role-based specifications of security protocols and operates in
the symbolic model. Roles are defined using state machines, similar to TLA 
\cite{Lamport:2002:SST:579617}, with high-level abstractions for cryptographic operations and message
passing.

\item \textit{ProVerif.} 
The ProVerif \cite{ProVerif_Source, BlanchetFOSAD14} specification language and
verification tool offer an extension of the pi-calculus \cite{Milner:1992:CMP:162037.162038} for role-based
specifications of security protocols in the symbolic security model. Each role
is defined as a process, and ProVerif offers significant resources for
specifying local computations as well as high-level abstractions for 
cryptographic operations and message passing between processes.

\item \textit{Tamarin.} 
The Tamarin \cite{TamarinSource, MSCB2013Tamarin} specification language and prover 
operate in the symbolic model, but does not
feature a role-based approach to specifying security protocols. Instead, the 
state of the protocol is represented as a multi-set (a bag) of facts, and 
protocol behavior is defined as rewrite rules that operate on this set of facts
(using, or consuming, old facts and producing new ones). 

\item \textit{CryptoVerif.} 
The CryptoVerif \cite{CryptoVerifSource, 4358700} specification language and 
prover offer a pi-calculus based language for role-based specifications of
security protocols, and it operates in the computational model~\cite{BlanchetTDSC07}. Abstractions 
for cryptographic operations are defined in terms of the security guarantees 
they provide. Proofs are defined as sequences of games \cite{Blanchet:2006:ASP:2165316.2165348}. 
In addition to proving the security properties, the CryptoVerif system can also 
transform specifications into OCaml implementations~\cite{6329273}.

\end{itemize}
}%end notes
\projname{} plus DistAlgo programs are simpler than even abstract specifications written in most of these formal specification languages.
%protocols are 
%nearly as concise as
%specifications written in Scyther.
Unlike these formal specification languages, \projname{} is for building actual implementations of security protocols as well as full-fledged secure applications.
\notes{
%CMK: 07/09/2018 - I definitely did not write the content of this.
Though they cannot be automatically verified,
using DistAlgo for implementing security protocols has the advantage that
DistAlgo has a formal operational semantics~\cite{liu2017clarity},
facilitating future automatic translation of DistAlgo specifications into other
formal protocol specification languages for verification.
}
%In addition, \projname{} can be used in implementing all security applications
%that require encryption-decryption and signing-verifying, not only security
%protocols.

In conclusion, %we have designed and implemented 
\projname{} provides simpler and more powerful high-level abstractions for
cryptographic operations and 
allows security protocols and applications to be written more easily and clearly.
%shown that they 
%can ease the development of secure
%applications and help prevent significant types of cryptographic misuse.
%CMK: I want to say something about writing clear and concise security
%protocols and secure applications, but not sure it fits.  
%
\notes{We are currently developing tools for automated translation of high-level
protocol implementations written using our abstractions to specification
languages like Scyther and ProVerif for automated verification.}
Future work includes
possible further optimization of the implementation to minimize performance overhead,
extension %of abstraction and implementation framework 
to support more combinations
of best cryptographic functions from different libraries,
static checking and optimization of these combinations,
%aannie: why the following?  i don't remember any reviewers asked this?
%CMK: This is the complaint that we have not demonstrated the "portability"
%CMK: of the abstractions.
%annie: i think it's ok/better to omit this---it doesn't sound technically strong
%CMK: Okay.
%implementation of the abstractions in additional programming languages,
more extensive use
and evaluation of the abstractions,
%conducting a user study to provide additional evidence for the effectiveness of the abstractions
and translation into languages of protocol verification tools such as ProVerif and 
Scyther for formal verification.

%Annie 5/6/18: this is a step too far from the current paper:
%automated generation of protocol implementations using our
%abstractions from simplified specifications similar to security protocol
%notation \cite{Briais:2007:FSP:1321786.1321979} (also known as Alice and Bob
%notation).
%CMK: 05/08/2018 - I agree.

\newcommand{\ackText}{
We thank Rahul Sihag for help in detailed running time measurements and analysis, and Yuege Chen and Wenjun Qu for protocol implementations in C\# and Java.  This work was supported in part by NSF under 
grants CCF-1414078 %distalgo 
and CNS-1421893 % scott-SaTC
and ONR under grant N000141512208.}

\forArxiv{\section*{Acknowledgements}
\ackText}

\forACM{
\begin{acks}
\ackText
\end{acks}
}

\appendix
% scott: I put this in a separate file, so it can easily be moved to Appendix for PLAS 2019 submission.
%annie: this section is not experimental, and not an evaluation either---need 
%       something specific to compare with, like other implementations in 
%       programming languages.
\section{Misuse prevention}
\label{sec:misuse-prevention}
%CMK: 06/12/2018 - Saksham raises the issue (which we have discussed
%CMK: before, about whether this is really something that can be reported as a result, 
%CMK: and suggests that it more properly belongs to the implementation section.

To validate SecAlgo against main types of cryptographic misuse, 
we surveyed studies of misuses that occur in mobile applications.

Table~\ref{tab-misuseTypesMerged} presents results of our evaluation 
using four such studies.  
It shows that SecAlgo prevents seven main types of cryptographic misuse out of a total of ten.
Note that each misuse type reported occurs at least once in each 
of the applications counted.  Thus even the sum
from CryptoLint alone means that SecAlgo abstractions prevent 
at least 13232 instances of types of cryptographic misuse.

\newcommand{\secalgoUnhandled}{*}
% A new table merging 1, 2 and 10
\newcommand{\captabMisuseTypesMerged}{Misuse type and number of apps containing that
type, plus total number of apps studied, using the misuse analysis systems 
CryptoLint~\cite{Egele:2013}, CMA~\cite{Shuai2014}, CNKX~\cite{Chatzikonstantinou:2016}, and CDRep~\cite{Ma:2016}.
Misuse types prevented by \projname{} are listed; three other types
studied~\cite{Egele:2013, Chatzikonstantinou:2016, Ma:2016} (a constant salt for
password-based encryption (PBE), $<$ 1000 iterations for PBE, and improper seeding for
Java SecureRandom objects), not prevented by \projname{}, are not listed.
'-' means that the study did not report about 
instances of the corresponding misuse type.
%Some types of cryptographic misuse and number of such misuses detected by each misuse
%analysis tool on Android applications. In %\cite{Egele:2013}, 11748 apps were analyzed
%using CryptoLint. In \cite{Shuai2014}, 45 apps were analyzed using CMA.
}
%\begin{table*}[htb]
\begin{table*}[htb]
\forSP{\renewcommand{\arraystretch}{1.3}}
\forSP{\caption{\captabMisuseTypesMerged}}
\centering
\resizebox{\textwidth}{!}{
%\begin{tabular}{@{}l|l||r|r|r|r@{}}%||r}
\begin{tabular}{@{~}l|l||r|r|r|@{~}r}%||r}
%\Xhline{2\arrayrulewidth}
\multirow{2}{*}{Misuse type} & \multirow{2}{*}{Description} & \multicolumn{4}{c}{Reported number of apps with this misuse type by study}\\
\cline{3-6}
& & CryptoLint \notes{\cite{Egele:2013}} & \hspace{5ex}CMA \notes{\cite{Shuai2014}} & \hspace{5ex}CNKX \notes{\cite{Chatzikonstantinou:2016}} & CDRep \notes{\cite{Ma:2016}} \\
%\Xhline{2\arrayrulewidth}
\hline\hline
M1K & Insufficient key size \notes{\cite{Shuai2014, Chatzikonstantinou:2016}} & - & 1 & 7 & - \\%
M2K & Constant or hardcoded keys \notes{\cite{Egele:2013, Shuai2014, Chatzikonstantinou:2016, Ma:2016}} & 3644 & 0 & 4 & 882 \\%
\hline
M1S & Encryption in ECB mode \notes{\cite{Egele:2013, Shuai2014, Chatzikonstantinou:2016, Ma:2016}} & 7656 & 7 & 16 & 887 \\%
M2S & Encryption with predictable IV \notes{\cite{Egele:2013, Shuai2014, Chatzikonstantinou:2016, Ma:2016}} & 1932 & 8 & 2 & 979 \\%
M3S & Encryption with obsolete algorithm \notes{\cite{Shuai2014, Chatzikonstantinou:2016}} & - & 8 & 16 & - \\%
\hline
M1A & RSA encryption without OAEP \notes{\cite{Shuai2014, Chatzikonstantinou:2016}} & - & 3 & 2 & - \\%
\hline
M1H & Hashing with obsolete algorithm \notes{\cite{Shuai2014, Chatzikonstantinou:2016, Ma:2016}} & - & 38 & 16 & 5582 \\%
\hline
Sum & Sum of numbers above & 13232 & 65 & 63 & 8330\\
\hline\hline
%Above & all types above & ... & ...\\
%All & all misuse types reported\\\hline\hline
%Apps with misuse & Number of apps analyzed with $\geq 1$ misuse & 10327 & 40 & 43 & 8582 \\%
%\hline
Total apps & Total number of apps analyzed & 11748 & 45 & 49 & 8640 \\%
%\Xhline{2\arrayrulewidth}
\hline
\end{tabular}
}%end \resizebox
\forArxiv{\caption{\captabMisuseTypesMerged}}
\forPrelim{\caption{\captabMisuseTypesMerged}}
\forACM{\caption{\captabMisuseTypesMerged}\vspacebelowfloat}
\label{tab-misuseTypesMerged}
\end{table*}

\newcommand{\capTabPrevent}{Summary of the way in which each misuse type is prevented by \projname{}.}
%\begin{table}[htb]
\begin{table}[htb]
\forSP{\renewcommand{\arraystretch}{1.3}}
\forSP{\caption{\capTabPrevent}}
\centering
%\resizebox{\columnwidth}{!}{
\begin{tabular}{@{~}l@{~}|@{~}p{.75\columnwidth}@{~}}
Misuse type & Prevention \\
\hline
\hline
M1K & excluded from whitelist of approved key sizes, safe defaults \\
\hline
M2K & \co{keygen} generates random key at runtime \\
\hline
M1S & excluded from whitelist of approved block modes of operation \\
\hline
%when called %annie: who calls
%CMK: 08/15/2018 - Need a short way to say that when encrypt is called to
%CMK: perform a block cipher encryption in a mode that requires an IV, SecAlgo
%CMK: generates the IV during the call to encrypt.
M2S & \co{encrypt} generates random IV when needed \\ 
\hline
M3S & excluded from whitelists of approved encryption algorithms \\
\hline
M1A & \co{encrypt} uses OAEP padding for RSA encryption, no alternative \\
\hline
M1H & excluded from whitelist of approved hasing algorithms \\
\hline
\end{tabular}
%}
\forArxiv{\caption{\capTabPrevent}}
\forPrelim{\caption{\capTabPrevent}}
\forACM{\caption{\capTabPrevent}}
\label{tab-prevent}
\end{table}

% OLD COMMENTED-OUT MATERIAL THAT USED TO BE HERE is now at end of this file.

We describe how exactly \projname{} prevents all of the misuse types 
listed in Table~\ref{tab-misuseTypesMerged}.  
They are summarized in Table~\ref{tab-prevent}.

%
%annie: inconsistent with said before; and in any case, do not need to repeat what's said before.
%\projname{} accounts for 7 out of the 11 misuse types found in the
%CryptoLint~\cite{Egele:2013}, CMA~\cite{Shuai2014}, and CDRep~\cite{Ma:2016}
%studies. 
%
%The way in which \projname{} prevents each misuse type are decribed in more detail below.
\begin{itemize}
\setlength{\itemsep}{0ex}
    \item \textbf{M1K: Insufficient key size.} This issue is handled by the
    \co{keygen} abstraction. The default key sizes for all algorithms are safe
    as they guarantee at least 112 bits of security, which NIST has determined
    as the minimum security strength allowable until
    2030~\cite{NIST:SP:800-57-1}. A key size given explicitly at a call to
    \co{keygen} is checked against a whitelist for the algorithm and if the key
    size is found to be insufficient, \projname{} throws an exception.
    
    \item \textbf{M2K: Hard-coded keys.} Hard-coded keys are unsafe because they
    can be extracted by binary disassembly. \projname{} inhibits the use of
    hard-coded keys through easy generation of keys using \co{keygen} and, in a planned extension, easy secure storage of keys.
    
    \item \textbf{M1S: Encryption in ECB mode.} Creation of a key for ECB mode
    is prevented during keygen because ECB is not included in the the whitelist
    of allowed block cipher modes in \projname{}. The whitelist is checked again
    when the key is used preventing the use of keys whose tags have been
    manually altered in an attempt to encrypt with ECB mode. Any attempt to use
    unsafe block modes like ECB, detected by checking the whitelist of approved
    modes will be reported as an error at runtime.  
    
    \item \textbf{M2S: Encryption with predictable IV.} The default behaviour of
    \co{encrypt} generates IVs automatically, thus preventing predictable IVs.
    \projname{} uses a cryptographically strong random number generator to
    generate the random data block to use as the IV as directed by NIST SP
    800-38A~\cite{Dworkin:2001:SRB:2206250}.
    
    \item \textbf{M3S: Encryption with obsolete algorithm.} As for M1S,
    this misuse is prevented by having a whitelist of safe algorithms.
    Obsolete algorithms like DES, ARC2 and ARC4 stream ciphers are not allowed
    by \projname{} abstractions \co{keygen}, \co{encrypt}, and \co{decrypt}. Any
    attempt to use an obsolete algorithm will be reported as an error at
    runtime.
    
    \item \textbf{M1A: RSA encryption without OAEP.} Optimal Asymmetric
    Encryption Padding (OAEP) is the default padding scheme used with RSA by
    \projname{}. \projname{} does not offer any alternative to OAEP and
    thereby ensures safety.
    
    \item \textbf{M1H: Hashing with obsolete algorithm.} Unsafe hashing
    algorithms like MD2, MD4, MD5 and SHA-1, are not in the whitelist of allowed
    hashing algorithms in \projname{}. Any attempt to use an obsolete algorithm
    will be reported as an error at runtime, as for M1S.
    
\end{itemize}

\notes{
Here we demonstrate the efficacy of our abstractions at curbing cryptographic misuse.
We begin by by reporting the proportion of observed instances of possible cryptographic misuse 
that would not have occurred had developers been using our abstractions.
}
\notes{
Table~\ref{tab-misuseTypesMerged} shows the numbers for misuse instances observed
in the studies from Section~\ref{sec-motive}, and they would all be prevented by 
the use of \projname{}. Altogether, use of \projname{} abstractions would prevent
about 64\% of the types of misuses looked for by the CryptoLint~\cite{Egele:2013}, 
CMA~\cite{Shuai2014}, and Mallodroid~\cite{Fahl:2012} misuse analysis tools. 
}
\notes{
%CMK: 05/15/2018 - It turns out this calculation makes no sense. Saksham was just
%CMK: rewriting a claim I had made in this section. This sloppiness is my fault.
Altogether, use of our abstractions would prevent 91\% of the 14371 found
instances of misuse.
}
\notes{The exceptions are the SSL/TLS certificate validation 
misuses found using the MalloDroid tool, which our current set of abstractions do not address.}
\notes{
\begin{table*}[h]
\centering
\begin{tabular}{l|c|c|c|c|c|c|c|c|c|c}
Misuse Analysis Tool                             & App Env.  & \# of Apps & M1K & M2K   & M1S   & M2S   & M3S & M1A   & M2A    & M1H \\
\hline
MalloDroid \cite{Fahl:2012}  & Android   & 6214       & -  & -    & -    & -    & -  &      & 1074  & -  \\
\hline
CryptoLint \cite{Egele:2013} & Android   & 11748      & -  & \cellcolor{lightgray} 3644 & \cellcolor{lightgray} 7656 & \cellcolor{lightgray} 1932 & -  & -    & -     & -  \\
\hline
CMA \cite{Shuai2014} & Android   & 45 & \cellcolor{lightgray} 1  & - & \cellcolor{lightgray} 7 & \cellcolor{lightgray} 8 & \cellcolor{lightgray} 8 & \cellcolor{lightgray} 3 & - & \cellcolor{lightgray} 38 \\
\hline
\end{tabular}
\caption{Shaded cells denote those misuses listed in Table~\ref{tab-misuseCount} count that would have been prevented by the use of \projname.}
\label{tab-misusePrev}
\end{table*}
}
\notes{
%CMK: 07/09/2018 - repetitive, delete
Misuses occur due to two factors: (1) the large number of decisions that must be made to call
cryptographic functions through APIs, and (2) the knowledge required to make those decisions
properly.
}
\notes{
%CMK: 07/09/2018 - Moving this passage to Section 5
Consider DS Simp protocol in Section~\ref{sec-appl}. It
contains three calls to \co{keygen}, two calls each to \co{encrypt} and \co{decrypt}, and one call each to
\co{sign} and \co{verify}.
}
\notes{
%CMK: 07/09/2018 - Moving this material to Section 4.
At the call to \co{keygen}, the programmer must decide whether to generate
a shared key or a public key, then decide whether this key will be used for encryption or signing. 
After choosing a kind of encryption and an operation, the programmer must then decide on an
algorithm. After choosing the algorithm, the programmer must then decide on a key size.

%CMK - TODO - Add title to draw attention to discussion of choices

From Table~\ref{tab-config}, we can count the options available at each subsequent decision.
For example, if the programmer selected shared key encryption, then there are three algorithms to select from,
and then three modes of operation. Each mode requires the generation of an auxilliary value. One of
them (CBC) requires that the plaintext is padded.

In total there can be between three and six significant decisions, and 22 possible outcomes at each call to
\co{keygen}. 

We address these opportunities for misuse by (1) denying access to unsafe choices, 
(2) automating decisions to deny programmers the opportunity to make unsafe choices, and 
(3) providing safe default selections for most of these decisions.
}
\notes{We address these opportunities for misuse in three ways. (1) We deny access to unsafe choices--weak algorithms,
vulnerable modes of operation, inadequate key sizes. This addresses the second cause of misuse by preventing
programmers from making bad choices due to ignorance. (2) We automate decisions to deny programmers the
opportunity to make bad choices. The generation of auxiliary values or the application of padding hidden
behind the abstraction. This addresses the first cause of misuse by reducing the number of significant 
decisions programmer are forced to make. (3) Finally, the provision of safe default selections for most
of these decisions obviates the need to make them at all. A programmer can choose a key type and know that
\projname{} will make safe choices for algorithm, key size, mode of operation, etc. by default.
}
\notes{Our efforts to reduce both the number and the danger of decisions required for access cryptographic
operations explains why so many instances of observed misuse would have been prevented by the use of
the \projname{} cryptographic interface.}

% OLD COMMENTED-OUT MATERIAL THAT USED TO BE IMMEDIATELY BELOW THE MISUSE TABLE
% I moved it here to get it out of the way.  --scott

\notes{
%CMK: 05/12/2018 - Made slight changes to table, not sure they are correct so I
%CMK: preserved the unchanged table here.
\newcommand{\secalgoUnhandled}{*}
% A new table merging 1, 2 and 10
\begin{table*}[htbp]
\centering
\begin{tabular}{l|l||r|r}
%\Xhline{2\arrayrulewidth}
\hline
\multirow{2}{*}{Misuse type} & \multirow{2}{*}{Description} & \multicolumn{2}{c}{Number of apps having this misuse type}\\
\cline{3-4}
& & CryptoLint \cite{Egele:2013} & CMA \cite{Shuai2014} \\
%\Xhline{2\arrayrulewidth}
\hline\hline
M1K & Insufficient key size \cite{Lazar:2014, Shuai2014} & - & 1\\
M2K & Hard-coded keys \cite{Egele:2013, Li2014, Lazar:2014, Shuai2014} & 3644 & -\\
\hline
M1S & Encryption in ECB mode \cite{Egele:2013, Li2014, Lazar:2014, Shuai2014} & 7656 & 7\\
M2S & Encryption with predictable IV \cite{Egele:2013, Li2014, Shuai2014} & 1932 & 8\\
M3S & Encryption with obsolete algorithm \cite{Lazar:2014, Shuai2014} & - & 8\\
\hline
M1A & RSA encryption without OAEP \cite{Shuai2014} & - & 3\\
\hline
 M1H & Hashing with obsolete algorithm \cite{Lazar:2014, Shuai2014} & - & 38\\
 \hline\hline
%Above & all types above & ... & ...\\
%All & all misuse types reported\\\hline\hline
Total Apps & Total number of apps analyzed & 11748 & 45\\
%\Xhline{2\arrayrulewidth}
\hline
\end{tabular}
\caption{Misuse type and number of apps containing that type, plus total number of apps studied, in \cite{Egele:2013} and \cite{Shuai2014}.
%Some types of cryptographic misuse and number of such misuses detected by each misuse analysis tool on Android applications. In %\cite{Egele:2013}, 11748 apps were analyzed using CryptoLint. In \cite{Shuai2014}, 45 apps were analyzed using CMA.
}
\label{tab-misuseTypesMerged}
\end{table*}
}
\notes{
\newcommand{\secalgoUnhandled}{*}
% A new table merging 1, 2 and 10
\begin{table*}[htbp]
\centering
\begin{tabular}{l|l|l|l|l}
\Xhline{2\arrayrulewidth}
\multirow{2}{*}{Misuse} & \multirow{2}{*}{Description} & \multicolumn{3}{c}{Misuse Analysis Tool}\\
\cline{3-5}
& & MalloDroid \cite{Fahl:2012} & CryptoLint \cite{Egele:2013} & CMA \cite{Shuai2014} \\
\Xhline{2\arrayrulewidth}
M1K & Insufficient key size \cite{Lazar:2014, Shuai2014} & - & - & 1\\
M2K & Hard-coded keys \cite{Egele:2013, Li2014, Lazar:2014, Shuai2014} & - & 3644 & -\\
\hline
M1S & Encryption in ECB mode \cite{Egele:2013, Li2014, Lazar:2014, Shuai2014} & - & 7656 & 7\\
M2S & Encryption with predictable IV \cite{Egele:2013, Li2014, Shuai2014} & - & 1932 & 8\\
M3S & Encryption with obsolete algorithm \cite{Lazar:2014, Shuai2014} & - & - & 8\\
\hline
M1A & RSA encryption without OAEP \cite{Shuai2014} & - & - & 3\\
M2A\secalgoUnhandled{} & Improper certificate validation \cite{Fahl:2012, Lazar:2014, Shuai2014} & 1074 & - & - \\
\hline
 M1H & Hashing with obsolete algorithm \cite{Lazar:2014, Shuai2014} & - & - & 38\\
\Xhline{2\arrayrulewidth}
\end{tabular}
\caption{Types of cryptographic misuse and number of such misuses detected by each misuse analysis tool on Android applications. In \cite{Fahl:2012}, 6325 apps were analyzed using MalloDroid. In \cite{Egele:2013}, 11748 apps were analyzed using CryptoLint. In \cite{Shuai2014}, 45 apps were analyzed using CMA. \secalgoUnhandled{}Misuses that would not have been prevented with \projname{}}
\label{tab-misuseTypesMerged}
\end{table*}
}
\notes{
\begin{table}[htbp]
\centering
\resizebox{55ex}{!}{%
\begin{tabular}{l|l|c}
Misuse & Description & References \\
\hline
M1K & Insufficient key size & \cite{Lazar:2014, Shuai2014} \\
\hline
M2K & Hard-coded keys & \cite{Egele:2013, Li2014, Lazar:2014, Shuai2014} \\
\hline
M1S & Encryption in ECB mode & \cite{Egele:2013, Li2014, Lazar:2014, Shuai2014} \\
\hline
M2S & Encryption with predictable IV & \cite{Egele:2013, Li2014, Shuai2014} \\
\hline
M3S & Encryption with obsolete algorithm & \cite{Lazar:2014, Shuai2014} \\
\hline
M1A & RSA encryption without OAEP & \cite{Shuai2014} \\
\hline
M2A & Improper certificate validation & \cite{Fahl:2012, Lazar:2014, Shuai2014} \\
\hline
M1H & Hashing with obsolete algorithm & \cite{Lazar:2014, Shuai2014} \\
\hline
\end{tabular}}
\caption{Types of cryptographic misuse with references.}
\label{tab-misuseTypes}
\end{table}
}
\notes{
\begin{table*}[htbp]
\centering
\begin{tabular}{l|c|c|c|c|c|c|c|c|c|c}
Misuse Analysis Tool                             & App Env.  & \# of Apps & M1K & M2K   & M1S   & M2S   & M3S & M1A   & M2A    & M1H \\
\hline
MalloDroid \cite{Fahl:2012}  & Android   & 6325       & -  & -    & -    & -    & -  &      & 1074  & -  \\
\hline
CryptoLint \cite{Egele:2013} & Android   & 11748      & -  & 3644 & 7656 & 1932 & -  & -    & -     & -  \\
\hline
CMA \cite{Shuai2014}                             & Android   & 45         & 1  & -    & 7    & 8    & 8  & 3    & -     & 38 \\
\hline
\end{tabular}
\caption{For each misuse analysis tool, the app mobile environment,
         number of apps analyzed, and number of misuses detected for each type of misuse.}
\label{tab-misuseCount}
\end{table*}
}
\notes{
%CMK: 08/13/2018 - Third try at stating the misuse types we do not address
\projname{} does not attempt to prevent the following misuses:
\begin{itemize}
	\item improper SSL/TLS certificate validation~\cite{Fahl:2012}
	\item improper use of password-based key derivation algorithms~\cite{Egele:2013, Chatzikonstantinou:2016, Ma:2016}
	\item improper seeding of SecureRandom~\cite{Egele:2013, Chatzikonstantinou:2016, Ma:2016}
\end{itemize}
For this reason we do not list these misuse types in Table~\ref{tab-misuseTypesMerged}.}
\notes{
%CMK: 05/12/2018 - Added explanation of the misuses we do not address.
Table~\ref{tab-misuseTypesMerged} does not list, and \projname{} does not
prevent, misuses associated with SSL/TLS certificate validation studied in
\cite{Fahl:2012, Shuai2014} or those misuses of password-based key derivation
operations and static seeding of SecureRandom (a Java/Android specific problem)
studied in~\cite{Egele:2013, Chatzikonstantinou:2016, Ma:2016}.\\
}
\notes{
%CMK: 4/2018 - Abbreviated the reference to the rest of these results.
In 2012, Fahl et al. \cite{Fahl:2012} analyzed 13,500 Android apps using their 
tool MalloDroid and found that 17\% apps that used SSL/TLS code improperly validated SSL certificates.
In 2015, Shuai et al. \cite{Shuai2014} applied their Crypto Misuse Analyzer to 45 popular Android apps 
(most had more than one million downloads) and found that 40 (89\%) of them contained at least one 
instance of cryptographic misuse. The iCryptoTracer tool, developed by Li et al. \cite{Li2014}, found
that 64 of the 98 (65\%) iOS apps to which it was applied contained at least one instance of 
cryptographic misuse.
}
\notes{
%CMK: 04/2018 - After the last revision this section ended up with two examples,
%CMK: which are redundant. I have removed the first example, and rely on the
%CMK: example in Figure 1.
Consider an example. The developer of an application wants to 
protect the secrecy of messages transmitted through the application, where 
secrecy is defined as the common security property of indistinguishability under
chosen plaintext attack (IND-CPA) \cite[pp. 6-7]{Schneier:1995:ACP:212584}. 
This can be done through the use of a block-cipher in combination with a mode of
operation~\cite{Dworkin:2007}. The developer must now make several choices to achieve
correct use of this cryptographic primitive, including the choice of a block-cipher
algorithm, a key size (and associated decisions about key generation, storage, and
access), a mode of operation, a padding method (if required by the algorithm or mode). 
etc.

However, not all combinations of decisions will provide message secrecy, defined as
IND-CPA. If the developer chooses the Advanced Encryption Standard (AES)~\cite{NIST197}
block-cipher, but opts for the Electronic Code Book (ECB) mode of operation
\cite{Dworkin:2007}, then the desired security property will be violated. ECB mode
encryption is not IND-CPA secure because, for a given key, identical blocks of plaintext
will result in identical blocks of ciphertext~\cite[p. 73]{Egele:2013}.
The developer believes that the secrecy of messages has been protected through the
use of the block cipher, but because the requirements for the correct use of that
cryptographic primitive have been violated, the secrecy of those messages cannot be
guaranteed.
%CMK: 04/2018 - This example requires lengthy discussion, and I am not sure it
%CMK: is neccessary. I have taken it out for now until we decide whether we need
%CMK: it, and until I figure out how to shorten it.
We can illustrate with the example in Figure~\ref{fig-cryptoExample}, which
displays a simple use of the PyCrypto~\cite{PyCrypto} cryptographic library for
Python. We use the Advanced Encryption Standard (AES)~\cite{NIST197} block
cipher in Cipher Block Chaining (CBC)~\cite{Dworkin:2007} mode to encrypt a
short message.

% should probably wrap code blocks that are longer than a few lines in a \begin{figure}
%\begin{figure}[htbp]
\begin{figure}[!t]
\centering
\fbox{%
\parbox{55ex}{%
\begin{code}
    from Crypto import AES\\
    from Crypto import Random\\
    key = Random.new().read(32)\\
    iv = Random.new().read(AES.block\_length)\\
    cipher = AES.new(key, AES.MODE\_CBC, iv)\\
    msg = iv + cipher.encrypt(pad(b'secret'))
\end{code}}}
\caption{Simple example of encryption using AES in CBC mode using the PyCrypto library.}
\label{fig-cryptoExample}
\end{figure}

Even in this simple example there are several decisions one must make and each
is an opportunity to make a mistake, and a mistaken choice is likely if one does
not possess some significant understanding of the use of block ciphers for
encryption or is not sufficiently careful.

\begin{enumerate}
\item \textit{Mode Selection} \\
If the mode is not specified as AES.MODE\_CBC, PyCrypto will default to the
insecure~\cite[p. 73]{Egele:2013} ECB mode of
operation~\cite{Dworkin:2007}.

\item \textit{Initialization Vector (IV) Use} \\
CBC mode requires the correct generation and use of an IV. The IV is a
block-size set of bytes that will be used to randomize the encryption of the
first block of plaintext. NIST standards \cite{Dworkin:2001:SRB:2206250} require
that the IV be unpredictable, which can be satisfied by generating the IV
randomly. However, we know from the misuse survey results included in
Table~\ref{tab-misuseCount} that developers routinely use non-random IVs.

In addition, the IV must be shared with the recipient of the ciphertext so that
it can be properly decrypted\cite[p. 20]{Dworkin:2001:SRB:2206250}. Common
practice is to prepend the IV to the ciphertext before sending it, but
developers who do not know that the IV can be safely transmitted in the clear
may not realize that this is an option. Even for experts it can be easy to
forget to transmit the IV along.

\item \textit{Key Generation} \\
The key must be generated randomly, and must be of an appropriate length. 
% BO - I think you can make the case stronger here -- the key should be random 
% BO - with good sources of entropy, as pseudo-random keys can be guessed
Example uses given in documentation, including the PyCrypto documentation
\cite{PyCrypto}, often use static keys, and do not indicate strenuously enough
how dangerous this is. In addition, key generation must use a cryptographically-
strong psuedo random number generator to ensure that randomly generated values
cannot be predicted by adversaries~\cite{NIST:SP:800-57-1}.
\end{enumerate}
}

\small
%\forSP{\IEEEtriggeratref{75}}
\forSP{\bibliographystyle{IEEEtran}}
\forACM{\bibliographystyle{ACM-Reference-Format}}
\forArxiv{\bibliographystyle{alpha}}
\bibliography{cryptoAbs}

\end{document}